\documentclass{emulateapj}
\usepackage{graphicx}
\usepackage{amsmath}
\usepackage{verbatim}
\usepackage{longtable}
\shorttitle{Size Evolution of Lyman Alpha Emitters}
\shortauthors{Bond et al.}

\begin{document}

\title{EVOLUTION IN THE CONTINUUM MORPHOLOGICAL PROPERTIES OF L\lowercase{y}$\alpha$-EMITTING GALAXIES FROM $\lowercase{z}=3.1$ TO $\lowercase{z}=2.1$\footnote{Based on observations made with the NASA/ESA Hubble Space Telescope, and obtained from the Hubble Legacy Archive, which is a collaboration  between the Space Telescope Science Institute (STScI/NASA), the Space Telescope European Coordinating Facility (ST-ECF/ESA) and the Canadian Astronomy Data Centre (CADC/NRC/CSA). }}

\author{Nicholas A. Bond and Eric Gawiser}
\affil{Physics and Astronomy Department, Rutgers University, Piscataway, NJ 08854-8019, U.S.A.}
\author{Lucia Guaita}
\affil{Department of Astronomy, Stockholm University, AlbaNova Science Center, SE-106 91 Stockholm, Sweden}
\author{Nelson Padilla\altaffilmark{2}}
\affil{Departmento de Astronomia y Astrofisica, Pontificia Universidad Catolica de Chile, Santiago, Chile}
\author{Caryl Gronwall and Robin Ciardullo}
\affil{Department of Astronomy and Astrophysics, Pennsylvania State University, University Park, PA 16802, U.S.A.} 
\author{Kamson Lai}
\affil{University of California Observatoires/Lick Observatory, University of California, Santa Cruz, CA 95064, USA}

\altaffiltext{2}{Centro de Astro-Ingenieria, Pontificia Universidad Catolica de Chile, Santiago, Chile}

\begin{abstract}

  We present a rest-frame ultraviolet morphological analysis of $108$ $z\simeq2.1$ Lyman Alpha Emitters (LAEs) in the Extended Chandra Deep Field South (ECDF-S) and compare it to a similar sample of $171$ LAEs at $z\simeq3.1$.  Using {\it Hubble Space Telescope} ({\it HST}) images from the Galaxy Evolution from Morphology and SEDs survey, Great Observatories Origins Deep Survey, and Hubble Ultradeep Field, we measure size and photometric component distributions, where photometric components are defined as distinct clumps of UV-continuum emission.  At both redshifts, the majority of LAEs have observed half-light radii $\lesssim 2$~kpc, but the median half-light radius rises from $1.0$~kpc at $z=3.1$ to $1.4$~kpc at $z=2.1$.  A similar evolution is seen in the sizes of individual rest-UV components, but there is no evidence for evolution in the number of multi-component systems.  In the $z=2.1$ sample, we see clear correlations between the size of an LAE and other physical properties derived from its SED.  LAEs are found to be larger for galaxies with higher stellar mass, star formation rate, and dust obscuration, but there is no evidence for a trend between equivalent width and half-light radius at either redshift.  The presence of these correlations suggests that a wide range of objects are being selected by LAE surveys at $z\sim2$, including a significant fraction of objects for which a massive and moderately extended population of old stars underlies the young starburst giving rise to the Ly$\alpha$ emission. 

\end{abstract}

\keywords{cosmology: observations --- galaxies: formation -- galaxies: high-redshift -- galaxies: structure}

\vspace{0.4in}

\section{INTRODUCTION}

\begin{deluxetable*}{lcccccc}
\tablecaption{LAE Samples\label{tab:MainSamples}}
\tablewidth{0pt}
\tablehead{
\colhead{Name}
&\colhead{Size}
&\colhead{HST Coverage\tablenotemark{a}}
&\colhead{GEMS}
&\colhead{GOODS}
&\colhead{HUDF}
&\colhead{Reference}\\
&
&
&
&
&
}
\startdata
z2Guaita	&250	&193	&175 &32 &6 & 1  \\
\t z2EWComplete\tablenotemark{b}	&130	&108 &97	&19 &6 & 1 \\
z3Gronwall	&154	&116  &97 &29 &3& 2  \\
z3Ciardullo	&62\tablenotemark{c}	&55\tablenotemark{c} &47\tablenotemark{c}	&12\tablenotemark{c} &0\tablenotemark{c} &3 \\
\enddata
\tablenotetext{a}{Counts the total number of objects covered by at least one of GEMS, GOODS, or HUDF}
\tablenotetext{b}{z2EWcomplete is a subsample of z2Guaita}
\tablenotetext{c}{Counts only those objects that are not already counted in z3Gronwall}

\tablerefs{
(1) Guaita et al. 2010; (2) Gronwall et al. 2007; (3) Ciardullo et al. 2011}

\end{deluxetable*}

The majority of galaxies in the local universe lie on the Hubble sequence \citep{HubbleSeq}, a continuum that runs between red, passively-evolving galaxies with a compact spheroidal component and gas-rich, star-forming disks with exponential light profiles.  This morphological sequence is seen clearly out to intermediate redshifts ($z \sim 1 - 2$) \citep[e.g.,][]{Glazebrook95,vdB96,Griffiths96,Brinchmann98,Lilly98,Simard99,vD00,Stanford04,Ravindranath04}, beyond which the majority of galaxies appear clumpy and irregular \citep[e.g.,][]{Giavalisco96,Lowenthal97,Dickinson,vdB01,Papovich05,Conselice05,Pirzkal07} and are difficult to place into existing classification schemes.  At $2 \lesssim z \lesssim 3.5$, these galaxies, the majority of which are actively star-forming, have sizes ranging from $< 1$~kpc to $\sim5$~kpc, with the largest often exhibiting multiple photometric components \citep[e.g.,][]{Bouwens04,Ravindranath06,Oesch09}.

One of the best-studied classes of galaxies at high redshift is the Lyman Alpha Emitter (LAE).  At $z \sim 3 - 4$, LAEs are widely believed to be actively star-forming, low in both stellar and dark matter mass, and relatively dust-free \citep[e.g.,][]{CowieHu,Venemans05,Gawiser07}.  The continuum morphological properties of LAEs vary from object to object, but the majority are compact ($C>2.5$) and have sizes $\lesssim 1.5$~kpc 
\citep{Venemans05,Pirzkal07,Overzier08,Taniguchi09,BondLAE,GronwallMorph}.  Multi-component or clumpy LAEs make up $\sim 20 - 45$\% of the population and typically have morphologies that are qualitatively similar to other star-forming galaxies at high redshift.  The emission-line morphologies of LAEs are difficult to study because of the long exposure times required, but \citet{Bond10} and \citet{Finkelstein10} find them to be spatially compact ($r_e<1.5$~kpc) at $z=3.1$ and $z=4.4$, respectively.  There may be evidence of an additional extended and diffuse component of Ly$\alpha$ emission in LAEs at $z=2.3$ \citep{NilssonLAE} and $z=4.4$ \citep{Finkelstein10}, but \citet{Bond10} found no evidence for this component in LAEs at $z=3.1$, so the question remains open.  Using ground-based imaging, \citet{SteidelHalos} found extended Ly$\alpha$ emission out to $\sim 80$~kpc around a stack of LAEs identified by the Lyman break technique, but it is unclear to what extent these highly extended halos contribute to the emission-line morphologies of typical LAEs on kiloparsec scales.

The Multiwavelength Survey by Yale-Chile \citep[MUSYC,][]{MUSYC} has obtained multiwavelength imaging and spectroscopy of $1.2$~degree$^2$ of sky in four fields, including the Extended Chandra Deep Field-South (ECDF-S).  As part of this survey, \citet{GuaitaLAE} (hereafter, Gu10) used broadband and $3727$~\AA\ narrow-band imaging of the ECDF-S to identify a large, unbiased sample of LAEs at $z=2.1$.  The authors found the LAEs in this sample to be weakly clustered, with a bias factor, $b \sim 1.8$, that is similar to that expected from the progenitors of present-day $L^*$ galaxies.  An analysis of the broadband optical and infrared colors of a ``stacked'' LAE \citep{z2SED} suggests that their median stellar masses ($\sim 4 \times 10^8$~M$_{\sun}$) are similar to those of $z=3.1$ LAEs \citep{Gawiser07}, although there is a subset of IRAC-bright objects that is $\sim10$ times more massive, on average, and exhibits non-negligible amounts of dust reddening, $E({\rm B}-{\rm V}) \sim 0.4$ \citep{Lai08,Acquaviva11}.

In this paper, we study the rest-UV continuum morphologies of the Gu10 sample of LAEs and compare them to those seen in LAEs at $z=3.1$ \citep{BondLAE,GronwallMorph} using the samples of \citet{GronwallLAE} and a new sample from \citet{z2LF}.  We use images taken by the Advanced Camera for Surveys (ACS) and obtained as part of the Galaxy Evolution from Morphology and SEDs survey \citep[GEMS,][]{GEMS}, Great Observatories Origins Deep Survey \citep[GOODS,][]{GOODS}, and Hubble Ultradeep Field survey \citep[HUDF,][]{HUDF}.  In what follows, we fit each photometric component separately and give quantitative size measures for both the individual components and the LAE system as a whole.  

In \S~\ref{sec:data} and \ref{sec:method}, we summarize the data and describe the analysis techniques used in our comparative study.  In \S~\ref{sec:results}, we give the photometric properties of each $z=2.1$ LAE system and its components and compare our results to those found for $z=3.1$ LAEs.  Finally, in \S~\ref{sec:discussion}, we discuss the implications of our findings for the physical nature of LAEs as a function of redshift. Throughout this paper, we will assume a concordance cosmology with $H_0=71$~km~s$^{-1}$~Mpc$^{-1}$, $\Omega_{\rm m}=0.27$, and $\Omega_{\Lambda}=0.73$ \citep{WMAP}.  With these values, $1\arcsec = 7.75$~physical~kpc at $z=3.1$ and $1\arcsec = 8.43$~physical~kpc at $z=2.1$.

\section{DATA}
\label{sec:data}

We use three LAE samples in our analysis, including the equivalent-width-complete subsample of $z = 2.1$ LAEs identified by Gu10, the flux-limited sample of $z\simeq3.1$ LAEs of \citet[][z3Gronwall]{GronwallLAE}, and the flux-limited sample of $z\simeq3.1$ LAEs of \citet[][z3Ciardullo]{z2LF}, all in the ECDF-S.  The V$_{606}$-band cutouts and morphological properties of z3Gronwall are already published in \citet[][hereafter B09]{BondLAE}.  All samples are summarized in Table~\ref{tab:MainSamples} and described below.

The initial sample of $250$ $z=2.1$ LAEs (hereafter, z2Guaita) was selected to have a monochromatic flux of $F_{\mathrm{Ly}\alpha} > 2 \times 10^{-17}$~ergs~cm$^{-2}$~s$^{-1}$, and rest-frame Ly$\alpha$ equivalent width of EW~$\gtrsim20$~\AA\null.  For some analyses, the authors made a further cut on Ly$\alpha$ luminosity, $L_{\mathrm{Ly}\alpha}>1.3 \times 10^{42}$~erg~s$^{-1}$, in order to remove a bias in their sample against high-EW objects.  We use this ``equivalent-width complete'' subsample (hereafter, z2EWcomplete) of $130$ $z=2.1$ LAEs for the majority of our morphological analyses.  Excluding objects within $40$ pixels of the edge of an image and cutouts with clear image defects, there are $108$ $z=2.1$ LAEs in z2EWcomplete that are covered by {\it HST} broadband imaging surveys.

The two $z\simeq 3.1$ LAE samples used here were both selected to have monochromatic fluxes, $F_{\mathrm{Ly}\alpha} > 1.5 \times 10^{-17}$~ergs~cm$^{-2}$~s$^{-1}$, and rest-frame Ly$\alpha$ equivalent widths, EW~$\gtrsim20$~\AA\null.  The z3Ciardullo sample covers the same field as z3Gronwall, but identifies LAEs using a narrow-band filter shifted $\sim 20$~\AA\ to the red relative to the z3Gronwall filter.  There remains some overlap between the two filters, however, so $\sim 50$\%\ of the objects identified in z3Ciardullo also appear in z3Gronwall.  We find $55$ additional LAEs in z3Ciardullo that have {\it HST} coverage, complementing the $116$ $z\simeq3.1$ LAEs studied in B09.  Note that we also exclude objects in these samples that are within $2$\arcsec\ of any X-ray source in the expanded catalogs of \citet{Lehmer05}, \citet{Virani06}, and \citet{Luo08}.

\subsection{GEMS}
\label{subsec:gems}

Of the surveys used in our study, the GEMS survey covers the largest area, consisting of $\sim 800$~arcmin$^2$ of the ECDF-S\null\ and $63$ ACS pointings in the F606W (V$_{606}$) and F814W (z$_{814}$) filters.  The depth of GEMS is relatively uniform across the field, with a $5\sigma$ detection limit of $m_{\rm AB} = 28.3$ for V$_{606}$-band point sources.  Survey images were multidrizzled \citep{Multidrizzle} to a pixel scale of $30$~mas.  
The cutouts for the LAEs in the $z=2.1$ and $z=3.1$ samples are shown in Figures~\ref{fig:GEMSPanels} and \ref{fig:GEMSPanelsKO3} (see end of article), respectively.

\subsection{GOODS}
\label{subsec:goods}

The southern half of the GOODS survey is a subregion of the Extended Chandra Deep Field-South, observing $\sim 160$~arcmin$^2$ of sky.  The GOODS observations included {\it HST}/ACS imaging in the F435W (B$_{435}$), V$_{606}$, F775W (I$_{775}$), and z$_{850}$ filters.  Although the V$_{606}$ imaging depth varies across the GOODS area, a typical $5\sigma$ detection limit for point sources is $m_{\rm AB} = 28.8$.  As in GEMS, all images were multidrizzled to a pixel scale of $30$~mas.   
The cutouts for the LAEs in the $z=2.1$ and $z=3.1$ samples are shown in Figures~\ref{fig:GOODSPanels} and \ref{fig:GOODSPanelsKO3} (see end of article), respectively.

\subsection{HUDF}
\label{subsec:hudf}

The Hubble Ultra-Deep Field (HUDF) reaches a V$_{606}$-band $5\sigma$ point source depth of $m_{\rm AB} = 30.5$ and includes {\it HST}/ACS imaging in the B$_{435}$, V$_{606}$, I$_{775}$, and z$_{850}$ filters.  The survey covers $11$~arcmin$^2$ of sky and has a multidrizzled pixel scale of $30$~mas.  
The cutouts for the LAEs in z2EWcomplete are shown in Figure~\ref{fig:HUDFPanels} (see end of article).

\section{METHODOLOGY}
\label{sec:method}

\subsection{Visual classification}
\label{subsec:classify}

\begin{figure}[t]
\figurenum{6}
\plotone{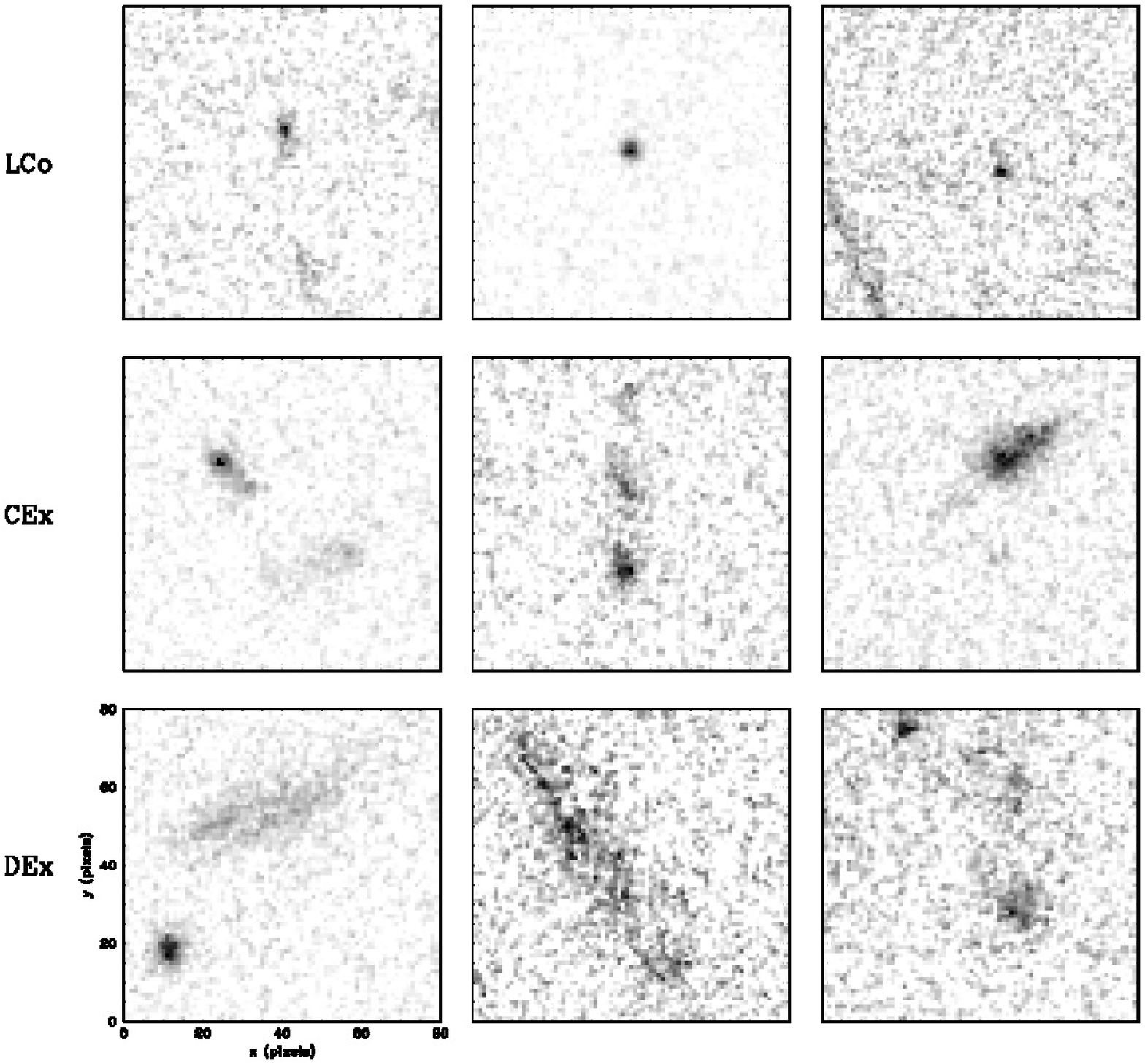}
\caption{Example cutouts of objects that are visually classified as diffuse-extended (DEx, bottom row), clumpy-extended (CEx, middle row), and compact (LCo, top row).  Galaxies in the DEx class are excluded from some of our analyses because of the high probability that they are coincident with low-redshift contaminants.  These cutouts are $2\farcs4$ on a side.
\label{fig:ClassPanels}}
\end{figure}

\begin{figure}[t]
\figurenum{7}
\plotone{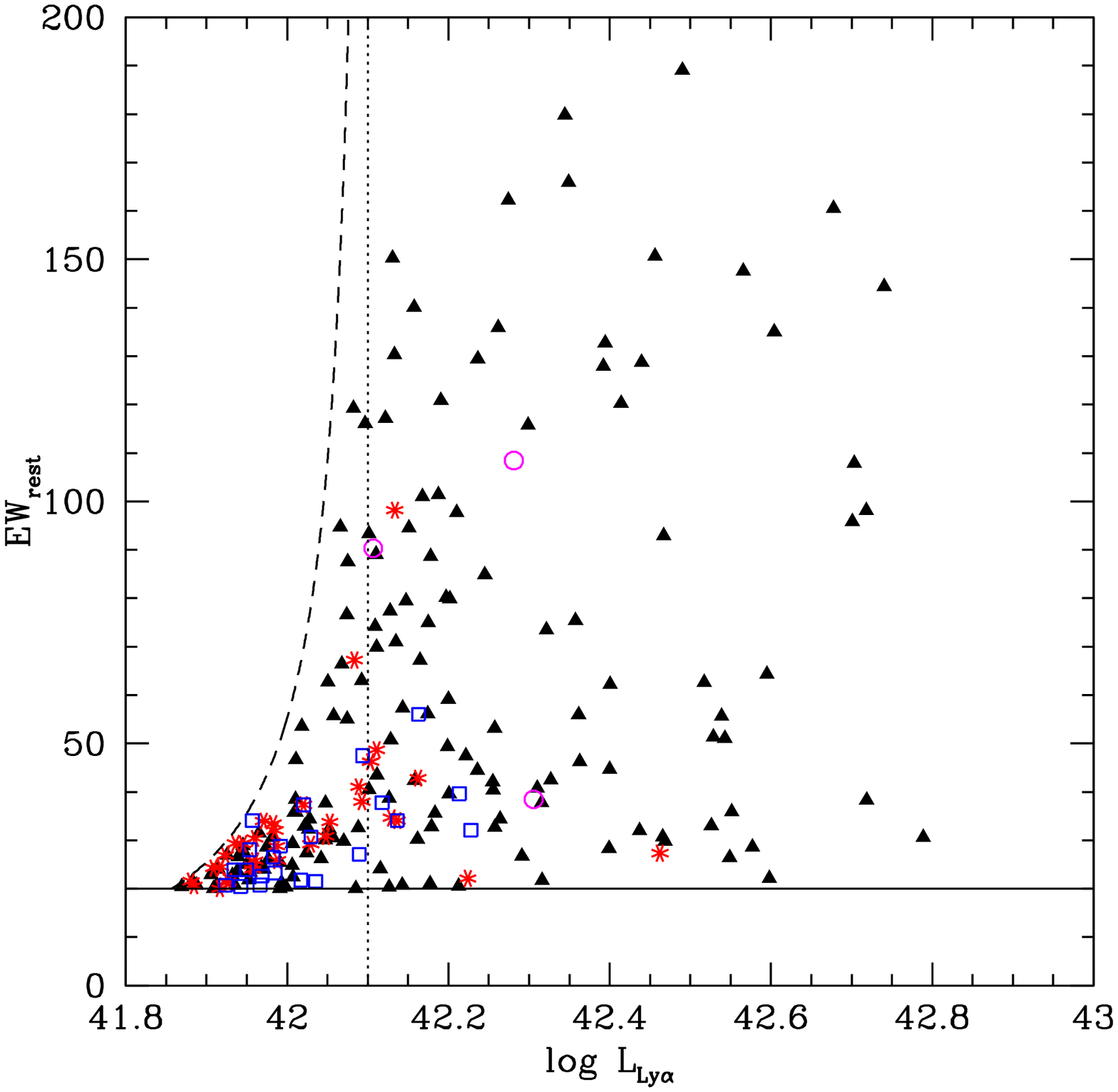}
\caption{Distribution of $z=2.1$ LAEs in rest-frame equivalent width vs. Ly$\alpha$ luminosity, with symbols indicating the visual morphological classification of the candidate LAE.  Objects with classifications of DEx, CEx, LCo, and NoD are plotted as stars, filled squares, filled triangles, and open circles, respectively.  The horizontal solid line is the $20$~\AA\ rest-frame equivalent width cutoff of the Gu10 survey and the dotted vertical line indicates the log$(L_{\mathrm{Ly}\alpha})>42.1$ luminosity cut used to make the z2EWcomplete sample.  The dashed curve delimits the selection region based on the $5\sigma$ limiting narrow-band magnitude of the Gu10 survey.
\label{fig:ContamClass}}
\end{figure}

We have classified all of the $193$ objects in z2Guaita with {\it HST} coverage according to their morphology in the V$_{606}$ cutouts.  The morphological classifications are determined by eye, taking into account an object's size and appearance, where the classes are chosen to separate objects with unusual physical conditions (e.g., apparent major mergers), as well as possible interlopers or contaminants.  The four classes are given below.
\begin{itemize}
\item LCo - Compact objects.  Much like the majority of LAEs at $z=3.1$, these objects are compact with little or no evidence for extended substructure.
\item CEx - Clumpy-extended objects.  They appear similar to the most extended high-redshift star-forming galaxies \citep[e.g., clump-clusters,][]{Elmegreen09}, with emission appearing in disconnected clumps.
\item DEx - Diffuse-extended objects.  These objects are large enough to be at least marginally resolved from the ground, with emission dominated by a diffuse component.  If they really are at $z=2.1$, they are very massive systems, but many are likely to be contaminants or low-redshift interlopers coincident with a high-redshift LAE.
\item NoD - Non-detected objects.  There is no discernable source in the V$_{606}$ cutout.
\end{itemize}
Examples of the LCo, CEx, and DEx classes are shown in Figure~\ref{fig:ClassPanels}. Of the objects in z2Guaita with {\it HST} coverage, a strong majority ($71$\%) are classified as LCo, with the CEx, DEx, and NoD classes making up $10$\%, $16$\%, and $3$\% of the sample, respectively.  

The candidate LAEs are plotted in equivalent width vs. Ly$\alpha$ luminosity in Figure~\ref{fig:ContamClass}, where symbol type indicates the visual classification.  Gu10 selected LAEs according to their narrow-band magnitude rather than their Ly$\alpha$ luminosity, leading to a bias against high-equivalent-width LAEs with ${\rm log}({\rm Ly}\alpha)<42.1$.  Therefore, the majority of our analyses are performed on the subsample with ${\rm log}({\rm Ly}\alpha)>42.1$, also referred to as z2EWcomplete.  The use of this subsample is further motivated by the small fraction ($6$\%) of objects in the DEx class with ${\rm log}({\rm Ly}\alpha)>42.1$, as these objects may be dominated by low-redshift contaminants or interlopers within the selection radius.

\subsection{Source extraction and aperture photometry}
\label{subsec:aperture}

In B09, we describe in detail a data analysis pipeline optimized for measuring the photometric properties of ``clumpy'' or irregular star-forming galaxies.  We will provide only a brief summary here.

For the z3Ciardullo sample, we followed B09 and extracted $80 \times 80$ pixel ($2\farcs4 \times 2\farcs4$) cutouts from the GEMS, GOODS, and HUDF images at the ground-based position of each LAE in our sample.  Our final sample includes only those LAEs with full survey coverage in the cutout region.  We identified all sources in each cutout using SExtractor \citep{SExtractor} with extraction parameters {\tt DETECT\_MINAREA}$=30$ and {\tt  DEBLEND\_MINCONT}$=0.06$, fitting and subtracting a uniform sky from each cutout.  In order to determine the rest-UV centroid of each LAE system, we again run SExtractor on each cutout, now with {\tt DETECT\_MINAREA}$=5$, and take the flux-weighted mean position of the resulting detections.  We follow the same procedure for the z2EWcomplete sample, with the exception that we extract somewhat larger cutouts,~$100 \times 100$ pixel ($3\arcsec \times 3\arcsec$), to provide full coverage for some of the more extended objects in the z2Guaita sample.

\begin{figure}[t]
\figurenum{8}
\plotone{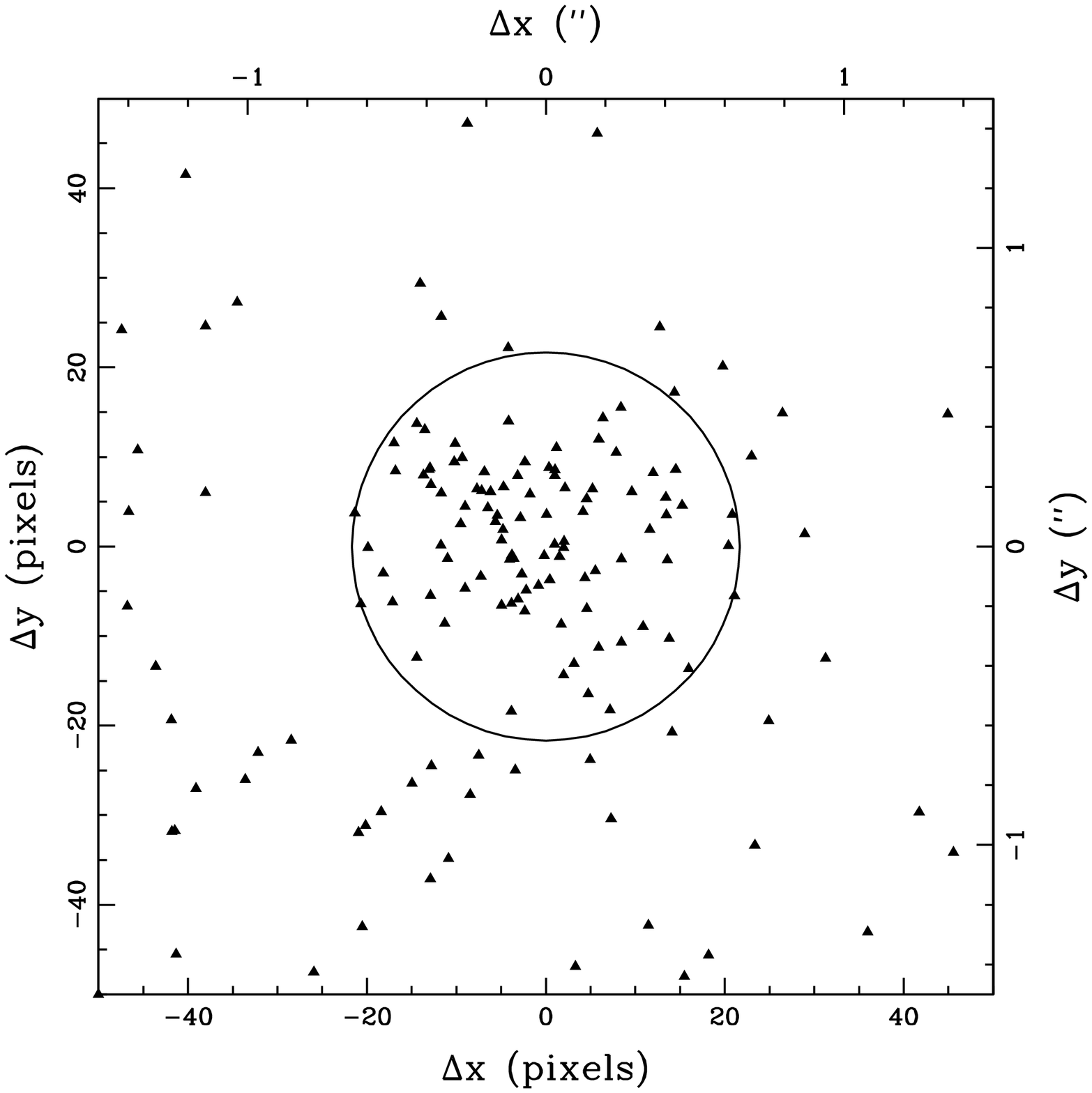}
\caption{Distribution of SExtractor detections in the V$_{606}$-band GEMS
cutouts as a function of distance from the ground-based
Ly$\alpha$ centroid.  We classify as LAE components any detections within the $0\farcs65$~($22$-pixel) selection
radius, drawn in black.
\label{fig:Exclusion_Radius}}
\end{figure}

For the z3Ciardullo sample, we follow B09 and select sources within $r_{\mathrm{sel}}\equiv0\farcs6$.  For z2EWcomplete, in order to account for the difference in the angular diameter distance between $z=3.1$ and $z=2.1$, we set $r_{\mathrm{sel}}\equiv0\farcs65$.  In Figure~\ref{fig:Exclusion_Radius}, we plot the distribution of SExtractor V$_{606}$-band detections as a function of angular distance from the ground-based Ly$\alpha$ centroid for the objects in z2EWcomplete.  Based on the density of sources outside the selection region, we estimate that at most $\sim 12$\%\ and $\sim 14$\%\ of all LAEs will have an interloper within $r_{\mathrm{sel}}$ in z3Ciardullo and z2EWcomplete, respectively.  The actual contamination rate will likely be less than this because the presence of an interloping source within $\sim 1$\arcsec\ will decrease the apparent equivalent width of the LAE in ground-based imaging and make it less likely to exceed the EW cutoff for the surveys.

Following B09, we define the photometric centroid to be the flux-weighted mean position of the detections within $r_{\mathrm{sel}}$.  Aperture photometry and half-light radii are computed within fixed apertures with radius, $r_{\mathrm{\mathrm{ap}}}=r_{\mathrm{sel}}$, using the IRAF routine {\tt PHOT} and centered on the centroid determined from the SExtractor runs.

\subsection{Monte Carlo simulations}
\label{subsec:Sims}

The {\it HST}/ACS images used in this study contain strong pixel-to-pixel correlations as a result of the drizzling process \citep{Multidrizzle}, so we estimate the uncertainties on the continuum magnitude and half-light radii using Monte Carlo simulations.  We perform a total of $10^5$ simulations for each survey, each time placing a Gaussian profile at a random position on the V$_{606}$-band image.  The simulated sources have Gaussian profiles with a dispersion between $2$ and $12$ pixels (between $0\farcs06$ and $0\farcs36$) and their photometric properties are computed within a fixed $0\farcs6$ aperture.  We define the scatter on a photometric measurement in terms of median statistics:
\begin{equation}
\sigma_x^m\equiv0.748[{\mathrm{Q3}}(x)-{\mathrm{Q1}}(x)],
\label{eq:medsig}
\end{equation}
where ${\mathrm{Q1}}$ and ${\mathrm{Q3}}$ are the first and third quartiles, respectively.  In the limit of a well-sampled Gaussian distribution, this quantity approaches the square root of the variance.

\begin{figure}[t]
\figurenum{9}
\plotone{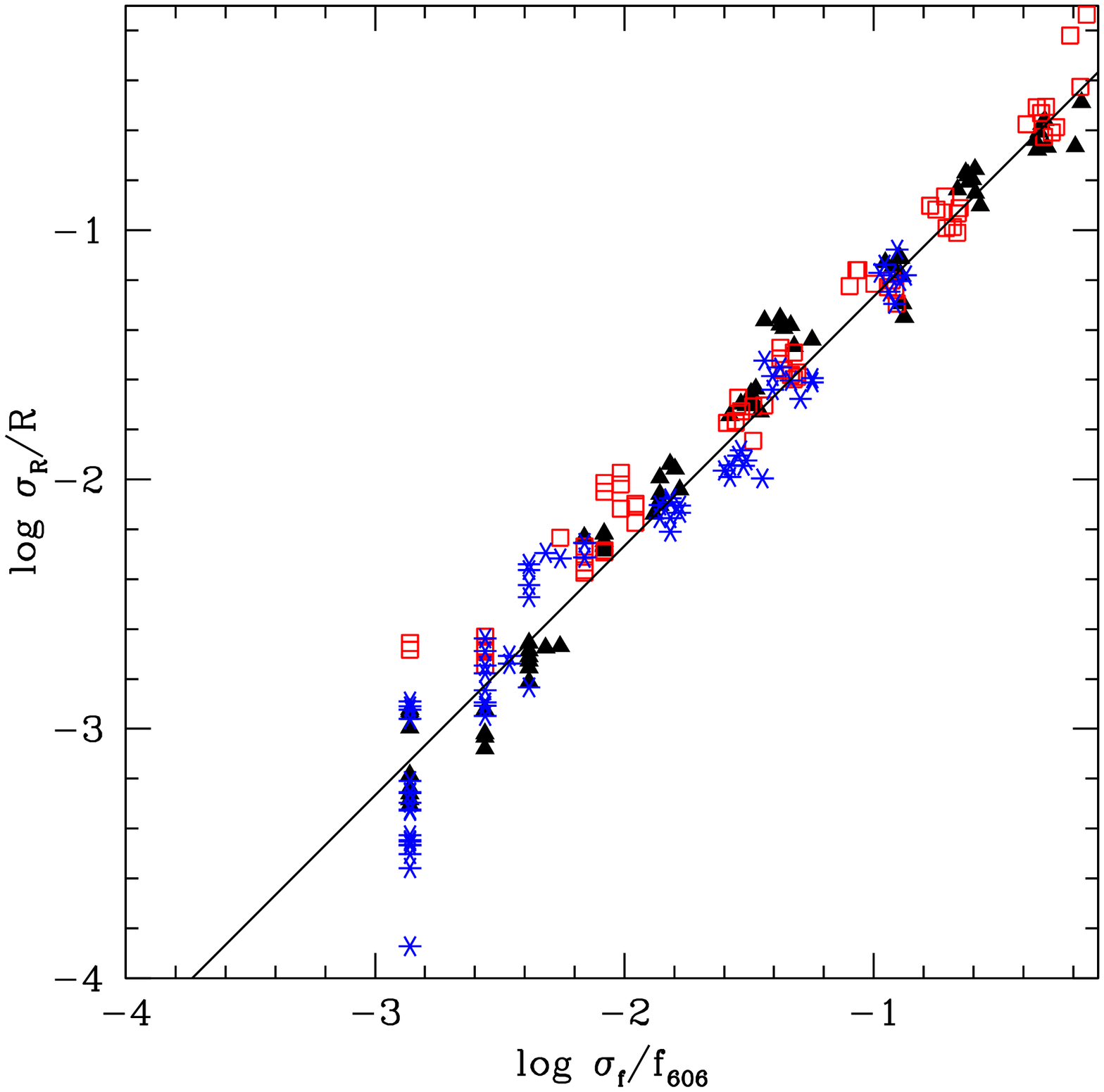}
\caption{Fractional uncertainties in the half-light radius as a function of the fractional uncertainty in flux through the V$_{606}$ filter for a
series of simulated Gaussian-shaped light profiles placed at random positions on the survey images.  Flux uncertainties are determined from the survey weight images and symbol type indicates the survey; filled triangles - GEMS, open squares - GOODS, stars - HUDF.  The solid line indicates the best-fit power law.
\label{fig:ReSims}}
\end{figure}

\begin{figure}[t]
\figurenum{10}
\plotone{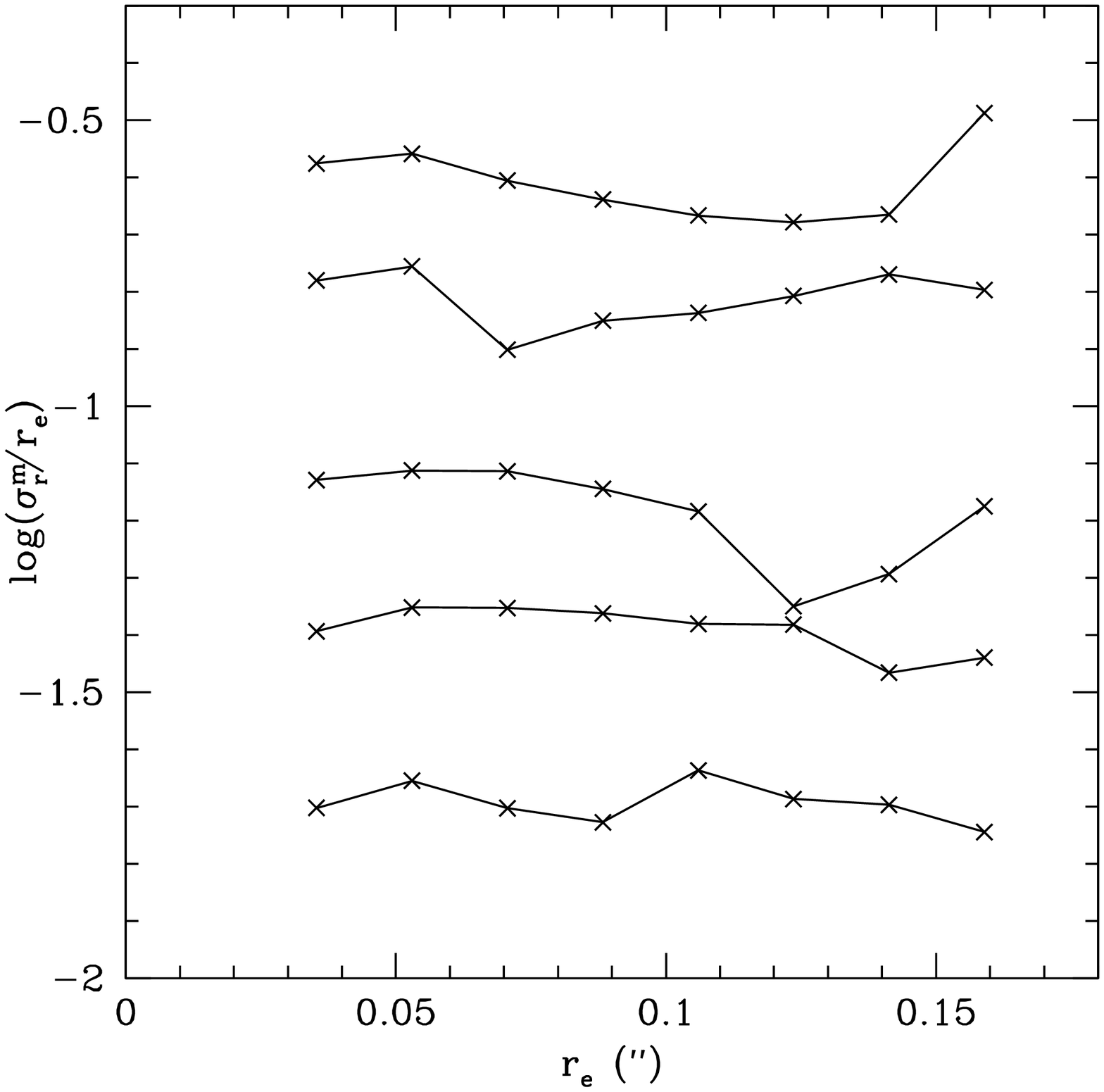}
\caption{Fractional uncertainty in the half-light radius as a function of half-light radius for a series of simulated Gaussian-shaped light profiles with constant continuum magnitudes.  Each curve represents a different continuum magnitude with, from top to bottom, V$_{606}=24.2$, $25.0$, $25.7$, $26.5$, and $27.2$.  Each cross symbol indicates the median over 1000 simulations.
\label{fig:ReSims_R}}
\end{figure}

The uncertainty on the fixed-aperture magnitudes can be estimated using the weight images provided with each survey, where the pixel-to-pixel flux uncertainty is given by $\sigma_f \equiv \overline{\sqrt{1/W}}$, with the mean being taken over pixel weight values, $W$, in each $0\farcs6$ aperture.  When we analyze the scatter in our simulations, we find a linear relationship between the fractional uncertainty in half-light radius and the fractional uncertainty in flux within the aperture (see Figure~\ref{fig:ReSims}).  For the Gaussian model profiles used in our simulations, a least-squares fit yields,
\begin{equation}
\frac{\sigma_{r_e}}{r_e}=0.54\frac{\sigma_f}{f_{606}}.
\end{equation}
As shown in the figure, this relationship holds for all of the surveys used here (GEMS, GOODS, and HUDF).  Furthermore, Figure~\ref{fig:ReSims_R} demonstrates that the fractional uncertainty in half-light radius is independent of half-light radius over the range typical for LAEs ($\sim 0\farcs06-0\farcs15$).

As noted in B09, there is a systematic tendency to overestimate the fixed-aperture half-light radius at very faint magnitudes because of the difficulties associated with measuring the light centroid (the center of the aperture) on a discrete grid.  This effect does appear in our half-light radius simulations, but it results in only a $\sim 4$\% overestimate of $r_e$ for marginally resolved sources with V$_{606}=27$.  The effect is even smaller for objects that are more extended and/or brighter than these limits.

\section{LAE MORPHOLOGY EVOLUTION BETWEEN $z=3.1$ AND $z=2.1$}
\label{sec:results}

Between the GEMS, GOODS, and HUDF surveys, there is {\it HST}/ACS coverage for a total of $171$ LAEs at $z=3.1$ and, considering only z2EWcomplete, $108$ LAEs at $z=2.1$.  When an object is covered by multiple surveys, we use the cutout from the deeper survey.  We note that there are two objects in z2EWcomplete (G$152$ and G$180$) ostensibly covered by the GOODS survey that have obvious defects in the V$_{606}$-band images.  Both of these objects also have GEMS coverage, so we instead analyze their defect-free GEMS cutouts. 

\subsection{UV continuum photometric centroid}
\label{subsec:centroid}

\begin{deluxetable}{lcccc}
\tablecaption{$z=2.1$ LAE Properties\label{tab:AllPhot}}
\tablewidth{0pt}
\tablehead{
\colhead{Number\tablenotemark{a}}
&\colhead{Survey}
&\colhead{$V^{\rm PHOT}$}
&\colhead{$d_c$\tablenotemark{b}}
&\colhead{$r_e^{\rm PHOT}$~\tablenotemark{c}}
\\
&
&\colhead{(AB mags)}
&\colhead{(\arcsec)}
&\colhead{(\arcsec)}
}
\startdata
G$75$ &HUDF &$27.07 \pm 0.04$ &$0.09$ &$0.15$  \\
G$76$ &HUDF &$28.21 \pm 0.10$ &$0.35$ &$0.10$  \\
G$78$ &HUDF &$26.09 \pm 0.01$ &$0.19$ &$0.13$  \\
G$83$ &HUDF &$26.23 \pm 0.01$ &$0.35$ &$0.14$  \\
G$90$ &HUDF &$27.04 \pm 0.03$ &$0.50$ &$0.14$  \\
\enddata

\tablenotetext{a}{Index}
\tablenotetext{b}{Distance between ACS and ground-based centroids}
\tablenotetext{c}{Half-light radius computed by {\tt PHOT} (not reported for LAEs without SExtractor detections)}
\tablenotetext{*}{This table is only a stub.  A manuscript with complete tables is available at http://www.nicholasbond.com/Bond0413.pdf}

\end{deluxetable}

The UV continuum emission in an LAE's host galaxy travels directly to us from the young stars, but Ly$\alpha$ nebular emission can resonantly scatter to large distances from the original starburst, depending on the distribution of neutral hydrogen surrounding the host galaxy.   In this circumstance, we might expect to see an offset between the emission-line centroid, measured in a narrow-band filter from the ground, and the rest-frame UV centroid, measured in the V$_{606}$ {\it HST}/ACS cutouts.  

Using the procedure described in Section~\ref{subsec:aperture}, we compute the V$_{606}$-band centroids of all of the objects with {\it HST} coverage.  There were five objects, one in z3Ciardullo (C34) and four in z2EWcomplete (G168, G206, G218, and G222), for which a centroid could not be computed because SExtractor did not detect any sources within $r_{\mathrm{sel}}$.  A visual inspection of their cutouts reveals no evidence for likely counterparts in C34, G206, or G222, but the remaining two have extended sources just outside $r_{\mathrm{sel}}$.  For the computation of fixed-aperture magnitudes and half-light radii, we set the centroids of these objects to their ground-based, narrow-band position (the cutout center).

\begin{deluxetable*}{lcccccc}
\tablecaption{$z=3.1$ LAE Properties\label{tab:AllPhot_KO3}}
\tablewidth{0pt}
\tablehead{
\colhead{Number\tablenotemark{a}}
&\colhead{Survey}
&\colhead{$\alpha$\tablenotemark{b}}
&\colhead{$\delta$\tablenotemark{b}}
&\colhead{$V^{\rm PHOT}$}
&\colhead{$d_c$\tablenotemark{c}}
&\colhead{$r_e^{\rm PHOT}$~\tablenotemark{d}}
\\
&
&
&
&\colhead{(AB mags)}
&\colhead{(\arcsec)}
&\colhead{(\arcsec)}
}
\startdata
C$15$ &GOODS &$3$:$33$:$03.233$ &$-27$:$50$:$48.260$ &$24.99 \pm 0.05$ &$0.49$ &$0.10$  \\
C$22$ &GOODS &$3$:$32$:$38.849$ &$-27$:$41$:$44.061$ &$24.80 \pm 0.03$ &$0.23$ &$0.20$  \\
C$34$ &GOODS &$3$:$32$:$17.457$ &$-27$:$42$:$45.900$ &$......$ &$...$ &$...$  \\
C$49$ &GOODS &$3$:$32$:$33.117$ &$-27$:$54$:$19.552$ &$26.32 \pm 0.08$ &$0.28$ &$0.23$  \\
C$52$ &GOODS &$3$:$32$:$54.834$ &$-27$:$46$:$40.132$ &$25.48 \pm 0.07$ &$0.14$ &$0.13$  \\
\enddata

\tablenotetext{a}{Index from table 2 of Ciardullo et al. 2010}
\tablenotetext{b}{Position of ACS centroid (set to ground-based position when there are no SExtractor detections)}
\tablenotetext{c}{Distance between ACS and ground-based centroids}
\tablenotetext{d}{Half-light radius computed by {\tt PHOT} (not reported for LAEs without SExtractor detections)}
\tablenotetext{*}{This table is only a stub.  A manuscript with complete tables is available at http://www.nicholasbond.com/Bond0413.pdf}

\end{deluxetable*}

\begin{figure}[t]
\figurenum{11}
\plotone{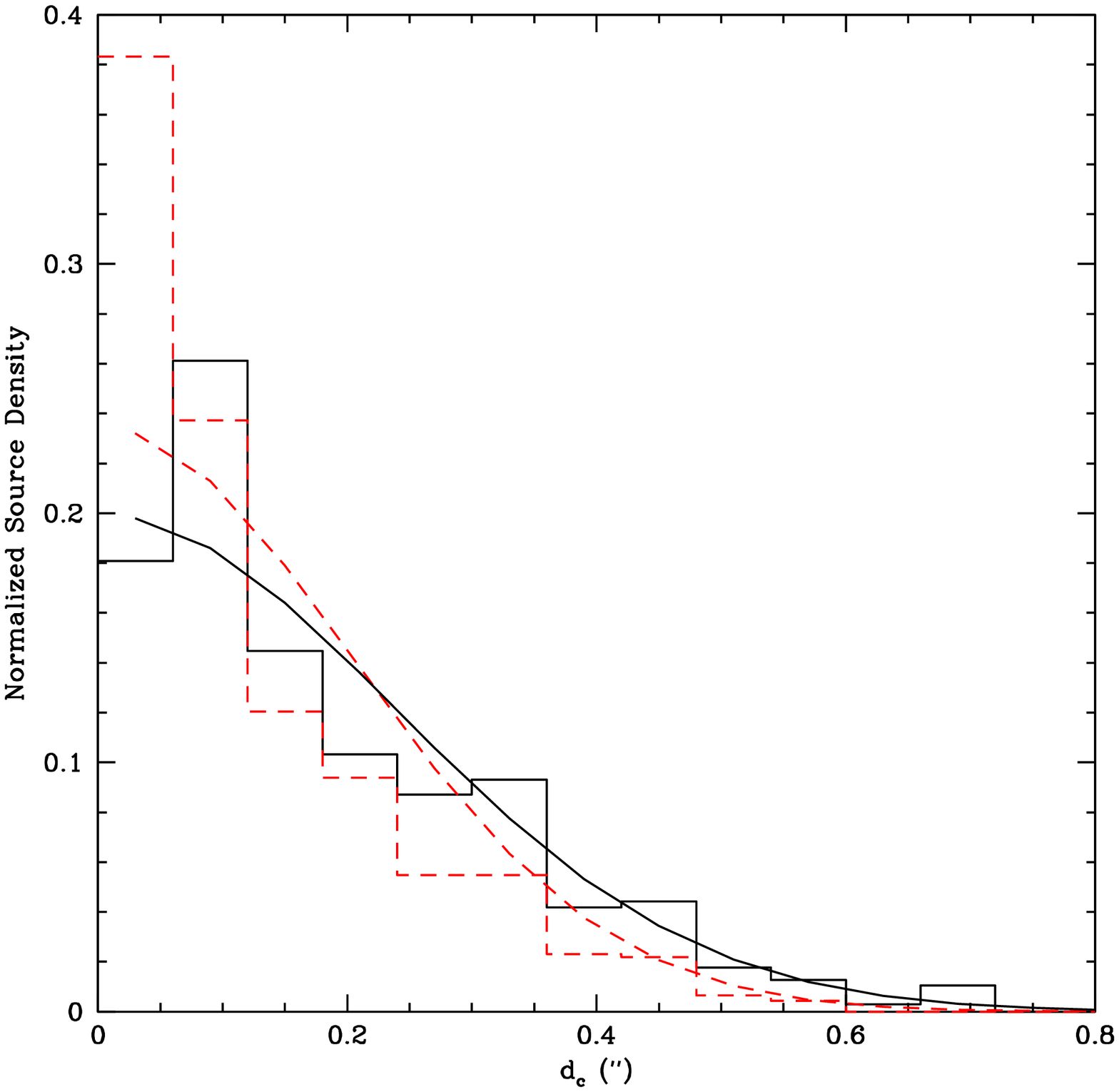}
\caption{Normalized distributions of distances between the V$_{606}$-band centroids and the ground-based narrow-band positions of LAEs at $z=2.1$ (solid histogram) and $z=3.1$ (dashed).  The frequency distribution for the best-fit two-dimensional Gaussians to each centroid offset distribution are also plotted as solid ($\sigma=0\farcs24$) and dashed ($\sigma=0\farcs20$) curves. The dispersions of these Guassians are consistent with the uncertainties in the ground-based astrometry ($0\farcs2-0\farcs3$).
\label{fig:CentroidFit}}
\end{figure}

In Figure~\ref{fig:CentroidFit}, we plot the distribution of measured offsets between the V$_{606}$-band continuum centroids and ground-based emission-line positions in both the z2EWcomplete catalog and the combined $z=3.1$ LAE catalog.  We fit two-dimensional Gaussians to the distributions of centroid offsets and find $\sigma=0.24 \pm 0.04$ and $\sigma=0.20 \pm 0.03$ for $z=2.1$ and $z=3.1$, respectively.  This amount of scatter is consistent with the expected $0\farcs2-0\farcs3$ astrometric uncertaintes in the ground-based narrow-band surveys, so we find no evidence for a physical offset between the emission-line and continuum light distributions at the $\sim 0\farcs2$ level ($\sim 1.5$~kpc at $z=2-3$) at either redshift.    

\subsection{Fixed-aperture photometric properties}
\label{subsec:fixedapz2}

The fixed-aperture half-light radius measures the size of the LAE ``system;'' that is, the size of the combined light distribution within $r_{\mathrm{sel}}$ of the rest-UV continuum centroid.  If, for example, an LAE originates from a pair of merging galaxies with comparable UV continuum brightness, then this half-light radius would be approximately the distance between the galaxies.  We give the fixed-aperture V$_{606}$ magnitudes, half-light radii, and centroid offsets (relative to the catalog position) for the objects in z2EWcomplete and z3Ciardullo in Tables \ref{tab:AllPhot} and \ref{tab:AllPhot_KO3}, respectively.  The corresponding tables for z3Gronwall can be found in B09.

\begin{figure}[t]
\figurenum{12}
\plotone{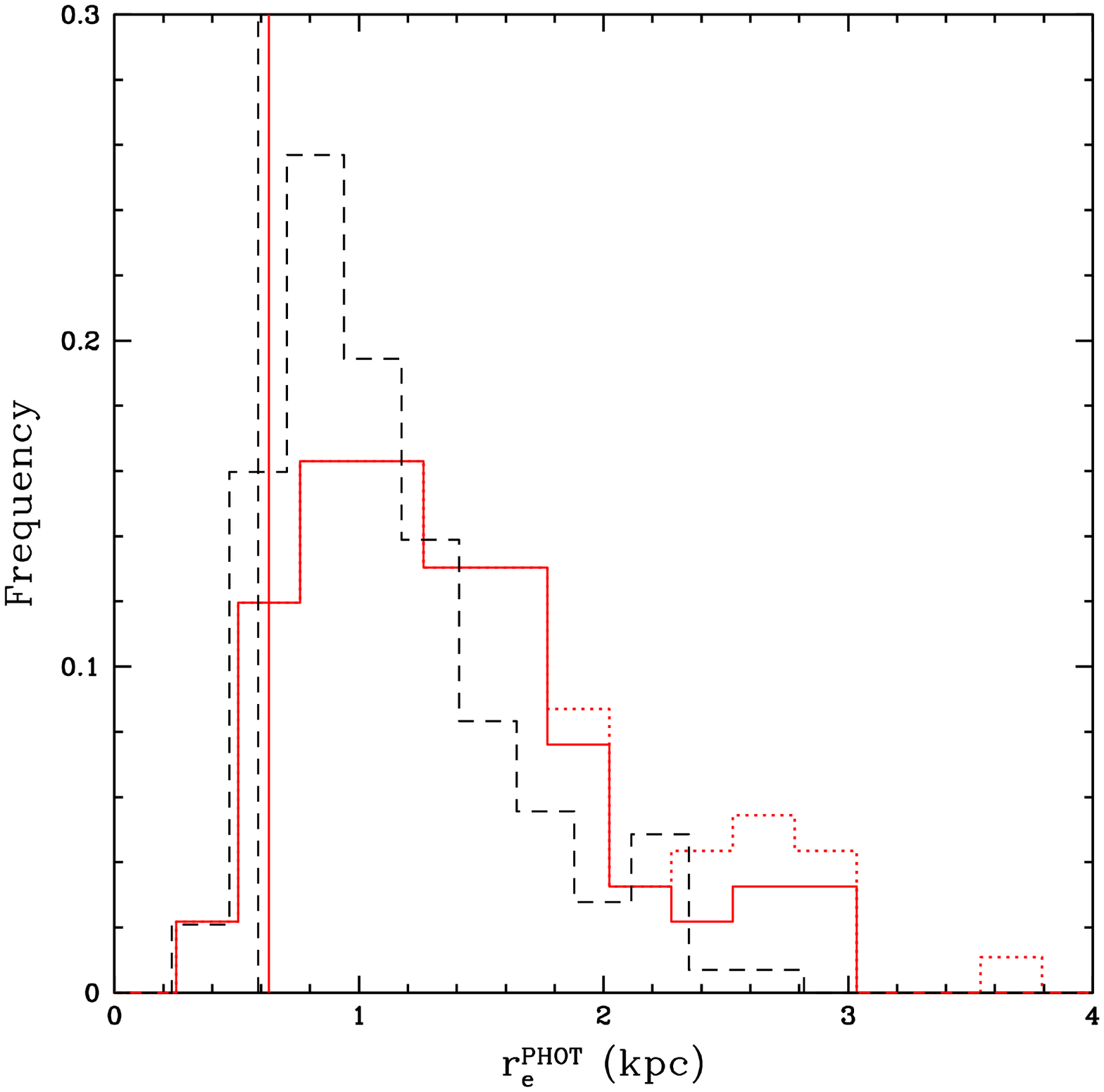}
\caption{Distributions of fixed-aperture, observed half-light radii for LAE systems at $z=2.1$ (solid and dotted) and $z=3.1$ (dashed).  The dotted histogram includes all objects in z2EWcomplete with {\it HST} coverage and the solid histogram excludes objects classified as DEx (see Section~\ref{subsec:classify}).  The vertical lines indicate the approximate resolution limit of the V-band {\it HST} images at each redshift.  We include only LAE systems with V$_{606}<28$. 
\label{fig:RePhotHist}}
\end{figure}

In Figure~\ref{fig:RePhotHist}, we plot the distribution of fixed-aperture half-light radii for z2EWcomplete and the combined $z=3.1$ LAE samples.  Where a large fraction of the $z=3.1$ LAEs are near the resolution limit, the majority $z=2.1$ LAE systems are resolved, with a tail extending to larger half-light radii.  The median half-light radii are $r_e^{\mathrm{PHOT}}=1.41$~kpc and $r_e^{\mathrm{PHOT}}=0.97$~kpc for z2EWcomplete and the combined $z=3.1$ sample, respectively.  Using the Kolmogorov-Smirnov (K-S) test and making the null hypothesis that the two samples are drawn from the same half-light radius distribution, we find a $P$-value of $0.0003$.  If we exclude those objects in z2EWcomplete classified as DEx in a visual inspection (and therefore less likely to be high-redshift LAEs, see Section~\ref{subsec:classify}), we find $P=0.004$.  Finally, even if we exclude all objects with large half-light radii, requiring $r_e^{\mathrm{PHOT}}<2.5$~kpc, we find $P=0.05$.  

We note that the V$_{606}$ filter is centered on rest-frame wavelengths of $\sim 1500$~\AA\ and $2000$~\AA\ at $z=3.1$ and $z=2.1$, respectively.  Although emission at both wavelengths is dominated by UV radiation from young stars, there may still be a difference in the apparent sizes of LAEs between these two parts of the spectrum.  To test for this, we compared the sizes of 18 $z=2.1$ LAEs in the observed-frame B- and V-band imaging from the GOODS survey.  We found the B-band (rest-frame $\sim 1500$~\AA) sizes to be $\sim 13$\% larger on average.  This may be due to diffuse Ly$\alpha$ emission leaking into the B-band filter, but it is difficult to tell from broadband imaging alone.  Regardless, these differences only act to increase the size evolution we measure between $z=2.1$ and $z=3.1$.  We therefore conclude that, for LAEs selected with the same rest-frame equivalent width and Ly$\alpha$ luminosity cutoffs, the systems at $z=2.1$ are systematically larger than those at $z=3.1$.

\begin{deluxetable*}{lcccccccc}
\tablecaption{$z=2.1$ LAE Component Properties\label{tab:Allsex}}
\tablewidth{0pt}
\tablehead{
\colhead{Number\tablenotemark{a}}
&\colhead{Component\tablenotemark{b}}
&\colhead{Survey}
&\colhead{$V^{\rm SE}$}
&\colhead{$d_c$\tablenotemark{c}}
&\colhead{$b/a$\tablenotemark{d}}
&\colhead{$\theta$\tablenotemark{e}}
&\colhead{$r_e^{\rm SE}$~\tablenotemark{f}}
&\colhead{SFR(UV)}
\\
&
&
&
&\colhead{(\arcsec)}
&\colhead{(AB mags)}
&\colhead{($^\circ$)}
&\colhead{(\arcsec)}
&\colhead{(M$_{\sun}$/yr)}
}
\startdata
G$75$ &$1$ &HUDF &$27.41 \pm 0.02$ &$0.03$ &$0.47$ &$-17.30$ &$0.14$ &$0.37$  \\
G$76$ &$1$ &HUDF &$28.29 \pm 0.04$ &$0.27$ &$0.61$ &$-11.40$ &$0.08$ &$0.16$  \\
G$78$ &$1$ &HUDF &$26.49 \pm 0.01$ &$0.17$ &$0.82$ &$51.80$ &$0.08$ &$0.85$  \\
$ $ &$2$ &HUDF &$27.42 \pm 0.02$ &$0.21$ &$0.70$ &$42.20$ &$0.09$ &$0.36$  \\
G$83$ &$1$ &HUDF &$26.50 \pm 0.01$ &$0.37$ &$0.93$ &$-46.60$ &$0.07$ &$0.85$  \\
\enddata

\tablenotetext{a}{Index from table 2 of Gronwall et al. 2007}
\tablenotetext{b}{Component number}
\tablenotetext{c}{Distance from ground-based Ly$\alpha$ position}
\tablenotetext{d}{Isophotal axis ratio computed by SExtractor}
\tablenotetext{e}{Isophotal position angle computed by SExtractor}
\tablenotetext{f}{Half-light radius computed by SExtractor}
\tablenotetext{*}{This table is only a stub.  A manuscript with complete tables is available at http://www.nicholasbond.com/Bond0413.pdf}

\end{deluxetable*}

\begin{figure}[t]
\figurenum{13}
\plotone{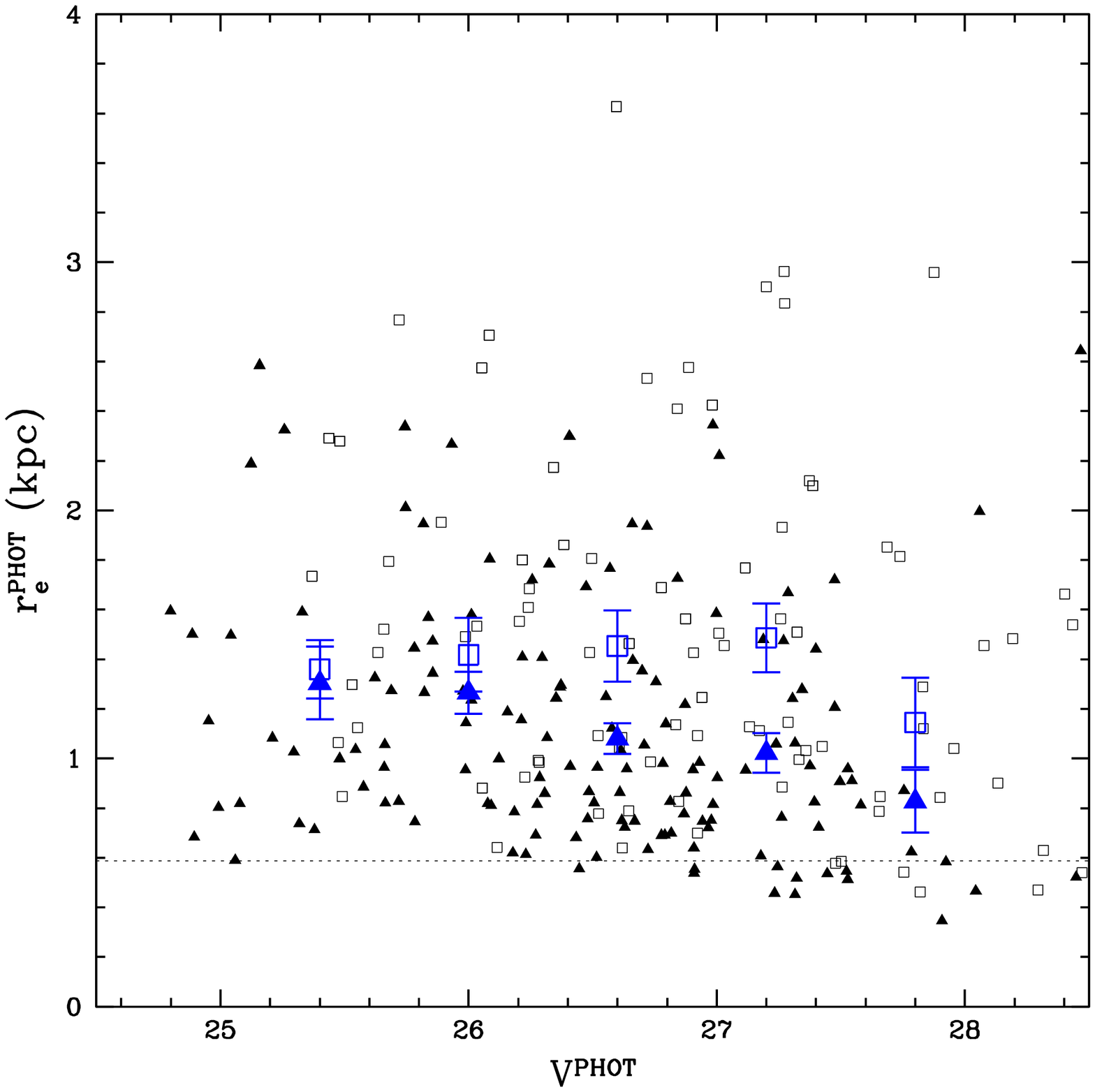}
\caption{Fixed-aperture, rest-UV half-light radius plotted versus
rest-UV continuum magnitude in the full
sample of LAEs with SExtractor detections, including objects at $z=3.1$ (solid triangles) and $z=2.1$ (open squares).  We have added $0.77$~mag to the $z=2.1$ V$_{606}$-band magnitudes, corresponding to the cosmological dimming that would occur if they were seen at $z=3.1$.  Small points indicate individual LAEs, while large points with error bars indicate the median half-light radii in bins of $\Delta {\rm V}_{606}=0.6$.  Uncertainties on the median are each computed from 5000 bootstrap simulations.  The dotted line
indicates the approximate resolution limit of the V$_{606}$-band {\it HST} images.
\label{fig:RpvV}}
\end{figure}

In Figure~\ref{fig:RpvV}, we show the dependence of fixed-aperture half-light radius on UV continuum magnitude for both z2EWcomplete and the combined $z=3.1$ samples.  To make the comparison more direct, we have added $0.77$~mag to the $z=2.1$ V$_{606}$-band magnitudes, corresponding to the cosmological dimming that would occur if they were seen at $z=3.1$.  At both redshifts, there is a great deal of scatter that is largely independent of continuum brightness and much larger than the observational uncertainties (which are $\sigma_r/r \sim 0.1$ at V$_{606}=26$, see Figure~\ref{fig:ReSims_R}).  In order to test for a correlation between UV continuum magnitude and half-light radius, we have divided the data into five bins in continuum magnitude, computed the mean half-light radius within each bin, and estimated the uncertainty on this mean using $5000$ bootstrap simulations.  The resulting means and their uncertainties are plotted as large points in Figure~\ref{fig:RpvV}.

There is no evidence for a correlation between $r_e^{\mathrm{PHOT}}$ and V$_{606}$ at $z=2.1$ - a constant value of $r_e=1.49$~kpc is a good fit at $\chi^2=2.6$.  However, the best-fit constant value to the combined $z=3.1$ LAE sample, $r_e^{\mathrm{PHOT}}=1.09$~kpc, is a significantly poorer fit to the data ($\chi^2=11.5$) than a model where size decreases logarithmically with continuum flux ($\chi^2=2.15$).  The best-fit two-parameter model to the $z=3.1$ sample is
\begin{equation}
r_e^{\mathrm{PHOT}}({\rm kpc})=-0.21 {\rm V}_{606}+28.8.
\end{equation}
It is worth noting that this relationship only deviates significantly from the $z=2.1$ data at faint continuum magnitudes, V$_{606}\gtrsim26.5$, suggesting that this is where most of the size evolution has occurred between $z=3.1$ and $z=2.1$.  

\begin{figure}[t]
\figurenum{14}
\plotone{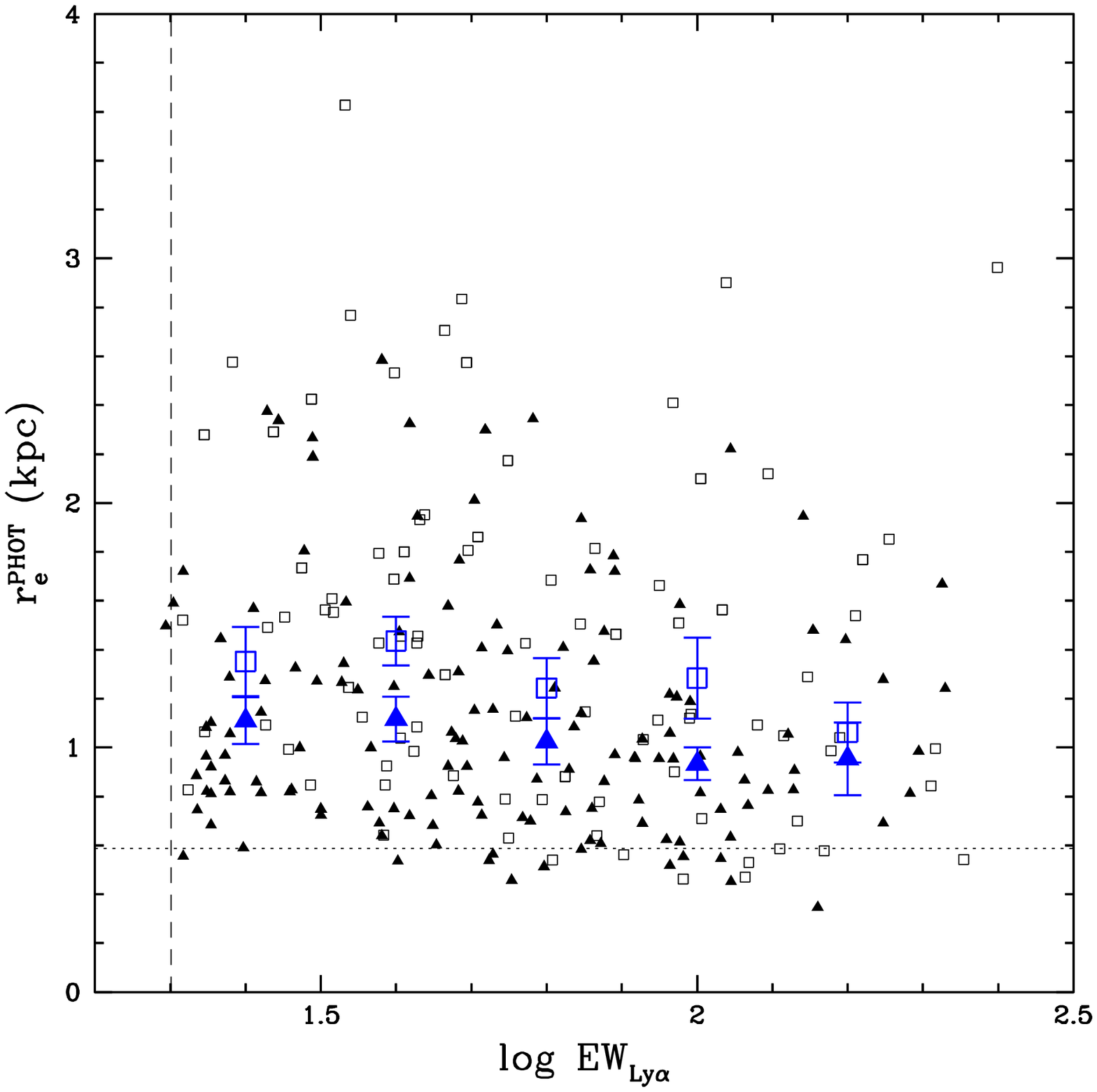}
\caption{ Same as Figure~\ref{fig:RpvV}, except plotting fixed-aperture, rest-UV half-light radius as a function of rest-frame Ly$\alpha$ equivalent width, where EW(Ly$\alpha$) measurements are taken from Gu10, \citet{z2LF}, and \citet{GronwallLAE}.  The dotted line indicates the approximate resolution limit and the dashed line indicates the approximate rest-frame EW limit of the LAE surveys.
\label{fig:RvEW}}
\end{figure}

We show a similar plot in Figure~\ref{fig:RvEW}, which gives the relationship between fixed-aperture half-light radius and rest-frame equivalent width.  We find no evidence for a correlation between equivalent width and half-light radius at either redshift.  This is in line with what is seen in a sample of Lyman break galaxies at $2.5<z<3.5$, which exhibits little dependence of the UV morphology on the presence or strength of Ly$\alpha$ emission \citep{Pentericci10}.

\subsection{Properties of photometric components}
\label{subsec:componentsz2}

\begin{deluxetable*}{lcccccccccc}
\tablecaption{$z=3.1$ LAE Component Properties\label{tab:Allsex_KO3}}
\tablewidth{0pt}
\tablehead{
\colhead{Number\tablenotemark{a}}
&\colhead{Component\tablenotemark{b}}
&\colhead{Survey}
&\colhead{$\alpha$}
&\colhead{$\delta$}
&\colhead{$V^{\rm SE}$}
&\colhead{$d_c$\tablenotemark{c}}
&\colhead{$b/a$\tablenotemark{d}}
&\colhead{$\theta$\tablenotemark{e}}
&\colhead{$r_e^{\rm SE}$~\tablenotemark{f}}
&\colhead{SFR(UV)}
\\
&
&
&
&
&
&\colhead{(\arcsec)}
&\colhead{(AB mags)}
&\colhead{($^\circ$)}
&\colhead{(\arcsec)}
&\colhead{(M$_{\sun}$/yr)}
}
\startdata
C$15$ &$1$ &GOODS &$3$:$33$:$03.232$ &$-27$:$50$:$48.267$ &$25.12 \pm 0.01$ &$0.44$ &$0.56$ &$-67.20$ &$0.10$ &$3.02$  \\
C$22$ &$1$ &GOODS &$3$:$32$:$38.850$ &$-27$:$41$:$44.056$ &$25.44 \pm 0.01$ &$0.22$ &$0.32$ &$-50.80$ &$0.20$ &$2.24$  \\
C$49$ &$1$ &GOODS &$3$:$32$:$33.108$ &$-27$:$54$:$19.636$ &$26.90 \pm 0.03$ &$0.21$ &$0.68$ &$-57.40$ &$0.10$ &$0.58$  \\
$ $ &$2$ &GOODS &$3$:$32$:$33.135$ &$-27$:$54$:$19.354$ &$27.73 \pm 0.07$ &$0.43$ &$0.83$ &$-21.90$ &$0.08$ &$0.27$  \\
C$52$ &$1$ &GOODS &$3$:$32$:$54.836$ &$-27$:$46$:$40.134$ &$25.75 \pm 0.02$ &$0.11$ &$0.82$ &$84.30$ &$0.11$ &$1.68$  \\
\enddata

\tablenotetext{a}{Index from table 2 of Ciardullo et al. 2010}
\tablenotetext{b}{Component number}
\tablenotetext{c}{Distance from ground-based Ly$\alpha$ position}
\tablenotetext{d}{Isophotal axis ratio computed by SExtractor}
\tablenotetext{e}{Isophotal position angle computed by SExtractor}
\tablenotetext{f}{Half-light radius computed by SExtractor}
\tablenotetext{*}{This table is only a stub.  A manuscript with complete tables is available at http://www.nicholasbond.com/Bond0413.pdf}

\end{deluxetable*}

In the terminology of this paper, a photometric ``component'' is any contiguous source within $r_{\mathrm{sel}}$ of the ground-based Ly$\alpha$ centroid of an LAE, where sources are identified in the V$_{606}$ {\it HST} images using SExtractor (see Section~\ref{subsec:aperture}).  We give the brightness, ellipticity, position angle and observed half-light radius ($r_e^{\rm SE}$) of each LAE component (as computed by SExtractor) for the z2EWcomplete and z3Ciardullo sample in Tables~\ref{tab:Allsex} and \ref{tab:Allsex_KO3}, respectively.  We quote half-light radii as computed by SExtractor rather than in fixed apertures in order to avoid blending effects in multi-component systems.

Photometric components can be galaxies themselves or individual star-forming regions within a larger galaxy -- it is difficult to distinguish these two possibilities from morphology alone for very compact objects like LAEs.  Furthermore, objects with multiple photometric components can be chance coincidences with low-redshift galaxies.  In z2EWcomplete, excluding objects classified as DEx, 19 of 108 objects in the sample have more than one photometric component.  This is consistent with the 23 of 171 objects ($13$\%) with multiple components at $z=3.1$, but it is also consistent with our estimated maximum rate of interlopers ($\sim 14$\%) in the selection region of the $z=2.1$ LAEs (based on the density of background sources in the cutouts, see Section~\ref{subsec:aperture}).  If the majority of multi-component systems in our LAE samples are contaminated by low-redshift interlopers, the distribution of component separations should be consistent with what we would see when comparing the LAE central components to a population with a random angular distribution.  To simulate this, we keep the brightest component in each multi-component system and randomly place another component within the selection circle.  

\begin{figure}[t]
\figurenum{15}
\includegraphics[scale=0.3,angle=270]{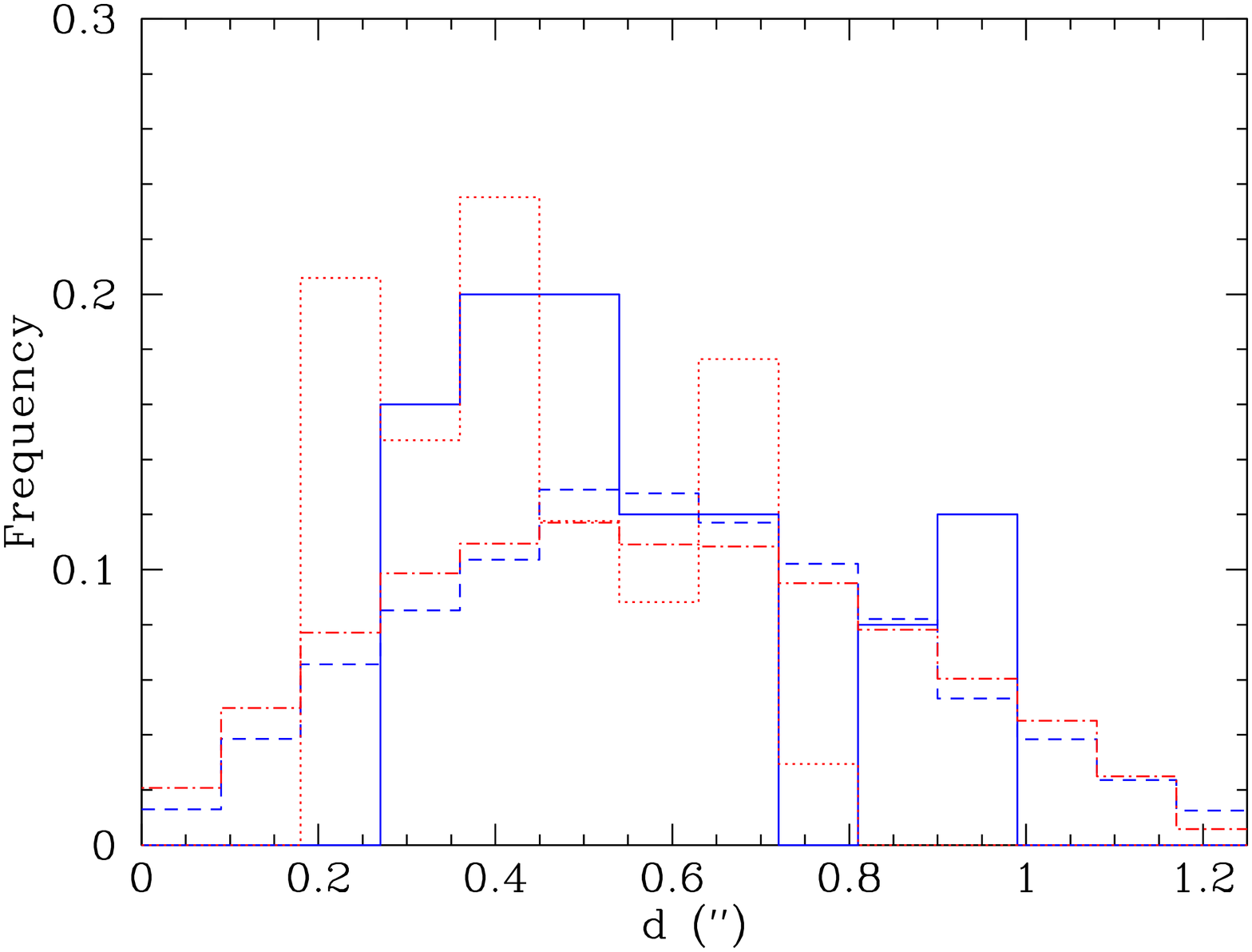}
\caption{Normalized distributions of separations between components in multi-component LAE systems.  When there are more than two components, we count as one every pair that includes the brightest component. The solid and dotted histograms are the distributions for the $z=2.1$ and $z=3.1$ LAE samples, respectively.  The dot-dashed and dashed histograms give the distributions we would get (for $z=2.1$ and $z=3.1$, respectively) if the fainter component of each pair were randomly distributed within the selection circle ($r_{\mathrm{sel}}=0\farcs65$ at $z=2.1$ and $0\farcs6$ at $z=3.1$).  
\label{fig:CompDists}}
\end{figure}

We show the normalized distribution of separations for the multi-component systems in our data and $10^4$ mock multi-component systems in Figure~\ref{fig:CompDists}.  At $z=3.1$, the real multi-component systems have a median separation of $0\farcs41$, while the same statistic is $0\farcs58$ for the mock systems.  With the null hypothesis that the two samples are drawn from the same parent distribution, the K-S test yields $P=0.029$, suggesting that the majority of the components in the multi-component systems in z3Ciardullo and z3Gronwall are associated with LAEs at $z=3.1$.  By contrast, we cannot exclude the null hypothesis for z2EWcomplete  ($P>0.1$), despite a smaller median separation in the observed systems ($0\farcs51$) than in the mock systems ($0\farcs58$).  Although this may mean that the multi-component systems in z2EWcomplete are heavily contaminated by interlopers, this result would also be consistent with an increase in the median separation between components in multi-component LAEs from $z=3.1$ to $z=2.1$.

\begin{figure}[t]
\figurenum{16}
\plotone{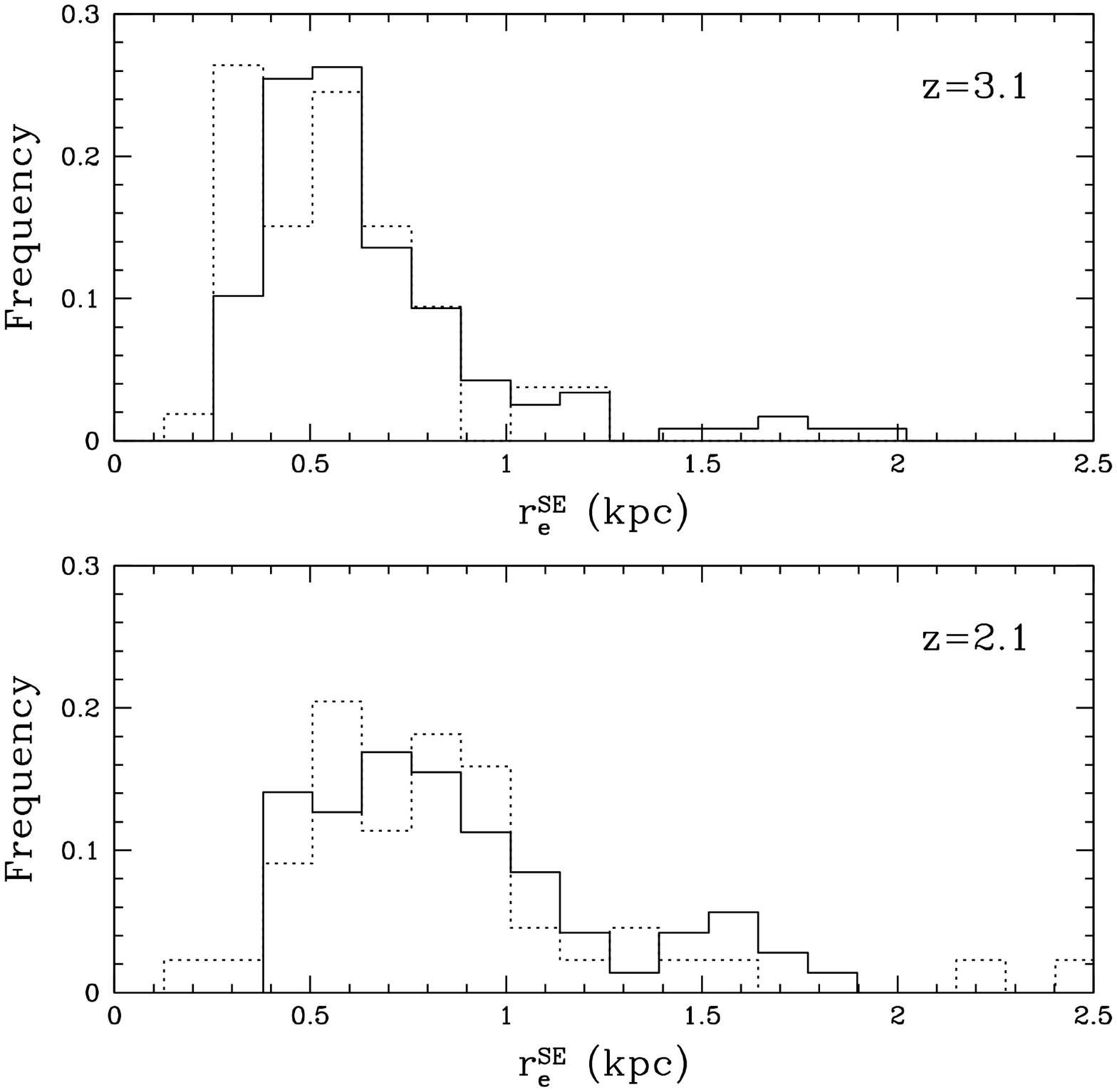}
\caption{Normalized SExtractor half-light radius distribution for components in single-component systems (solid) and multi-component systems (dotted) for the $z=3.1$ (top panel) and $z=2.1$ (bottom) LAE samples.   We only plot LAE components with V$_{606}<26.2$. 
\label{fig:CompHists}}
\end{figure}

Another way to discern the nature of the multi-component systems is to compare their component size distribution to that of single-component systems.  If low-redshift interlopers are common in the multi-component systems, these components might also have larger half-light radii than the average LAE.  We show the component size distributions in single- and multi-component systems in Figure~\ref{fig:CompHists}.  The K-S test reveals no evidence that the single- and multi-component systems are drawn from a different parent distribution at either $z=2.1$ or $z=3.1$ ($P>0.1$).  It is therefore unlikely that the multi-component systems at either redshift are dominated by low-redshift interlopers, unless their size distribution is very similar to that of high-redshift LAEs.

In B09, we showed that the half-light radius measured by SExtractor was systematically underestimated for components detected at S/N$\lesssim30$.  For the shallowest survey used here (GEMS), this corresponds to V$_{606}\simeq26.2$, so we construct subsamples of components brighter than this limit (including those in multi-component systems).  At $z=3.1$, the remaining components have $\mathrm {med}(r_e^{SE})=0.79$~kpc, compared to $\mathrm {med}(r_e^{SE})=0.95$~kpc at $z=2.1$.  With the null hypothesis that the two samples are drawn from the same parent distribution, the K-S test gives $P=0.07$, again consistent with an increase in the size of the typical LAE from $z=2.1$ to $z=3.1$.  

\subsection{Median morphological properties of LAE subsamples}
\label{subsec:medstat}

\begin{deluxetable*}{lcccc}
\tablecaption{Sizes of z=2.1 LAE Subsamples\label{tab:Subsamples}}
\tablewidth{0pt}
\tablehead{
\colhead{Subsample\tablenotemark{a}}
&\colhead{Selection}
&\colhead{Number of LAEs\tablenotemark{b}}
&\colhead{Median Size\tablenotemark{c}}
&\colhead{Size Spread\tablenotemark{d}}
\\
&
&
&\colhead{(kpc)}
&\colhead{(kpc)}
}
\startdata
UV-faint &$R>25.5$ &$55$ &$1.13 \pm 0.09$ &$0.53 \pm 0.16$  \\
UV-bright &$R<25.5$ &$78$ &$1.49 \pm 0.09$ &$0.84 \pm 0.13$  \\
IRAC-faint &$f_{3.6}<0.57$~$\mu$~Jy$$ &$38$ &$1.10 \pm 0.11$ &$0.66 \pm 0.13$  \\
IRAC-bright &$f_{3.6}>0.57$~$\mu$~Jy$$ &$30$ &$1.48 \pm 0.27$ &$1.24 \pm 0.18$  \\
low-EW &$EW<66$~\AA$$ &$39$ &$1.23 \pm 0.14$ &$0.69 \pm 0.12$  \\
high-EW &$EW>66$~\AA$$ &$31$ &$1.12 \pm 0.10$ &$0.64 \pm 0.21$  \\
red-LAE &$B-R>0.5$ &$19$ &$1.64 \pm 0.21$ &$0.83 \pm 0.23$  \\
blue-LAE &$B-R<0.5$ &$114$ &$1.27 \pm 0.10$ &$0.77 \pm 0.09$  \\
\enddata

\tablenotetext{a}{\,LAE subsamples defined in Guaita et al. 2010}
\tablenotetext{b}{Includes only LAEs with {\it HST}/ACS coverage}
\tablenotetext{c}{Median fixed-aperture half-light radius within $0\farcs65$}
\tablenotetext{d}{Interquartile range of fixed-aperture half-light radius}

\end{deluxetable*}

\citet{z2SED} selected a series of subsamples from the z2Guaita LAEs, separately dividing the sample by Ly$\alpha$ equivalent width, rest-frame UV luminosity, $3.6$~$\mu$m brightness, and B$-$R color.  We present the morphological properties of these subsamples in Table~\ref{tab:Subsamples}.  In addition to median half-light radius, we give the interquartile range (Q3-Q1) of the half-light radii.  Since measurement errors are subdominant to intrinsic scatter in the LAE sizes (see Section~\ref{subsec:fixedapz2}), the latter quantity will give an indication of the morphological heterogeneity within the subsample.  Uncertainties on each quantity are estimated from $10^4$ bootstrap simulations.

For comparison, we also create a set of subsamples at $z=3.1$.   We use the same  Ly$\alpha$ equivalent width selection criterion as at $z=2.1$, but correct the rest-frame UV luminosity and $3.6$~$\mu$m brightness selection criteria for the difference in luminosity distance.  Finally, for the rest-UV color, we use V$-$R instead of B$-$R, attempting to match the rest-frame wavelengths of the bluer band.  The properties of each subsample are shown in Table~\ref{tab:Subsamplesz3}. 

We find no significant size difference between the high- and low-equivalent width subsamples at either redshift, consistent with the object-by-object results shown in Figure~\ref{fig:RvEW}.  The UV-bright subsamples are larger in both size and size spread than the UV-faint subsamples at both redshifts, but the difference is greater and more statistically significant at $z=2.1$, with $\Delta r_e=0.36 \pm 0.13$.  A statistically significant difference only appears when we include all of z2Guaita - when the subsamples are restricted to objects in z2EWcomplete, the half-light radius difference between UV-bright and UV-faint subsamples decreases to $\Delta r_e=0.17 \pm 0.18$.

The other two pairs of subsamples, those selected by $3.6$~$\mu$m flux and B$-$R color, also show a difference between their median half-light radii at $z=2.1$.  In general, the subsample with the larger median size has properties consistent with what we would expect from objects with a dusty, older stellar population underlying the recent star formation.  The brightness at $3.6$~$\mu$m, in particular, is an indicator of the total stellar mass of an LAE and will be large when a galaxy has undergone previous bursts of star formation.  At $z=3.1$, we still see a significant size difference between the red and blue LAEs, but the fraction of red LAEs is smaller ($11$\% as compared to $17$\% at $z=2.1$), indicating that the dusty objects, although still larger in size at $z=3.1$, make up a smaller fraction of the LAE population.  We see no significant difference in size between the IRAC-faint and IRAC-bright subsamples at $z=3.1$, and the fraction of IRAC-bright objects is again smaller ($26$\%, as compared to $44$\%).

All of the subsamples exhibit a size spread, as indicated by the interquartile range of the half-light radius distribution, that is considerably larger than that expected from the measurement uncertainties alone.  The smallest spread ($0.5$~kpc) is seen in the UV-faint subsample, but its fractional spread (IQR[$r_e$]/med[$r_e$]) is comparable to that of the other subsamples.  The most noticeable outlier is the IRAC-bright subsample, which exhibits a size spread of $1.2$~kpc, suggesting an unusual amount of heterogeneity in this subsample.  A comparable spread is not seen in the $z=3.1$ IRAC-bright subsample.

\section{DISCUSSION}
\label{sec:discussion}

The case for evolution in the LAE population between $z\sim3$ and $z\sim2$ is becoming quite strong.  \citet{NilssonLAE}, comparing samples of LAEs at $z\sim 3$ and $z=2.25$, claim an increase in the AGN fraction and the UV-to-Ly$\alpha$ SFR ratio, as well as a narrower EW distribution at $z=2.25$.  Analyzing the sample studied in this paper, Gu10 confirmed an increase in UV-to-Ly$\alpha$ SFR toward $z\sim2$, but found no evidence for an increase in the AGN fraction at $z=2.1$.  Also using the Gu10 sample, \citet{z2LF} find a $\sim 50$\% decrease in the number density for sources with $L>1.5\times 10^{42}$~erg~s$^{-1}$, as well as a decrease in both the scale length of the equivalent width distribution and the characteristic luminosity ($L^*$) of the luminosity function at $z=2.1$.  

\begin{figure}[t]
\figurenum{17}
\plotone{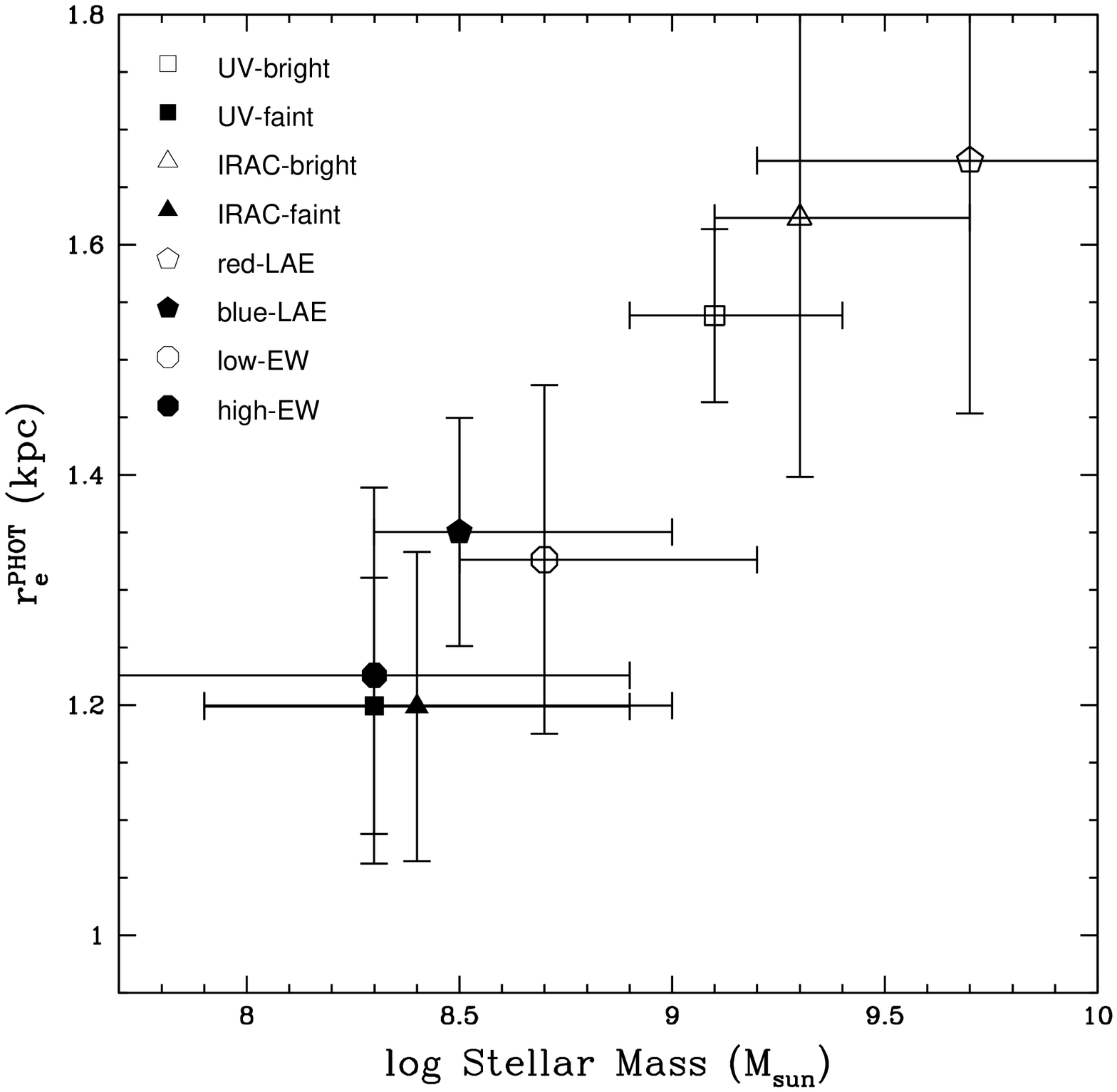}
\caption{Median fixed-aperture half-light radius versus stellar mass in subsamples of $z=2.1$ LAEs.  Stellar masses are taken from an SED median stacking analysis in Guaita et al. 2010.  
\label{fig:RvMass}}
\end{figure}

\begin{figure}[t]
\figurenum{18}
\plotone{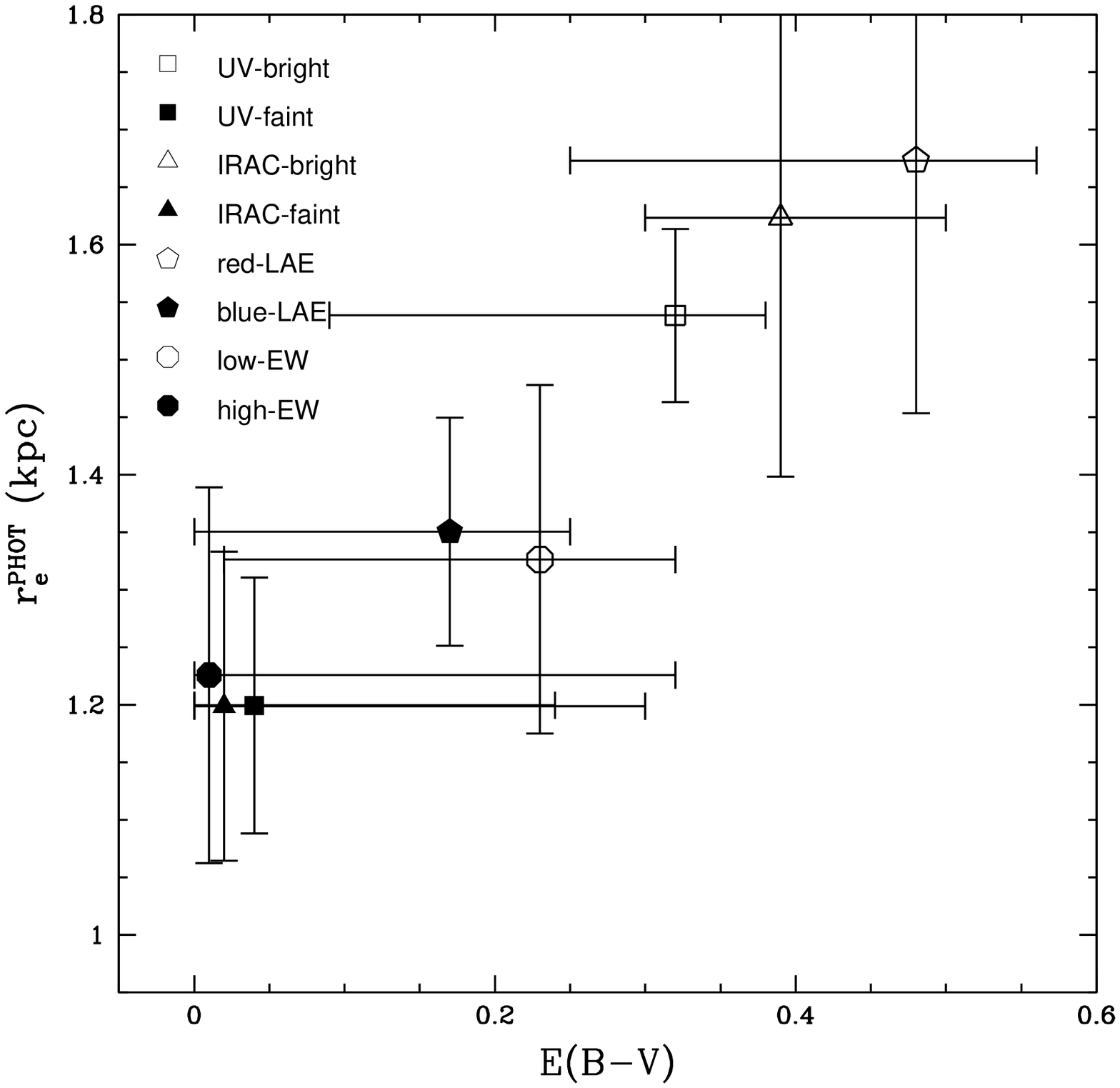}
\caption{Same as Figure~\ref{fig:RvMass}, but plotting fixed-aperture half-light radius as a function of dust reddening, $E(\mathrm{B}-\mathrm{V})$.
\label{fig:RvDust}}
\end{figure}

\begin{figure}[t]
\figurenum{19}
\plotone{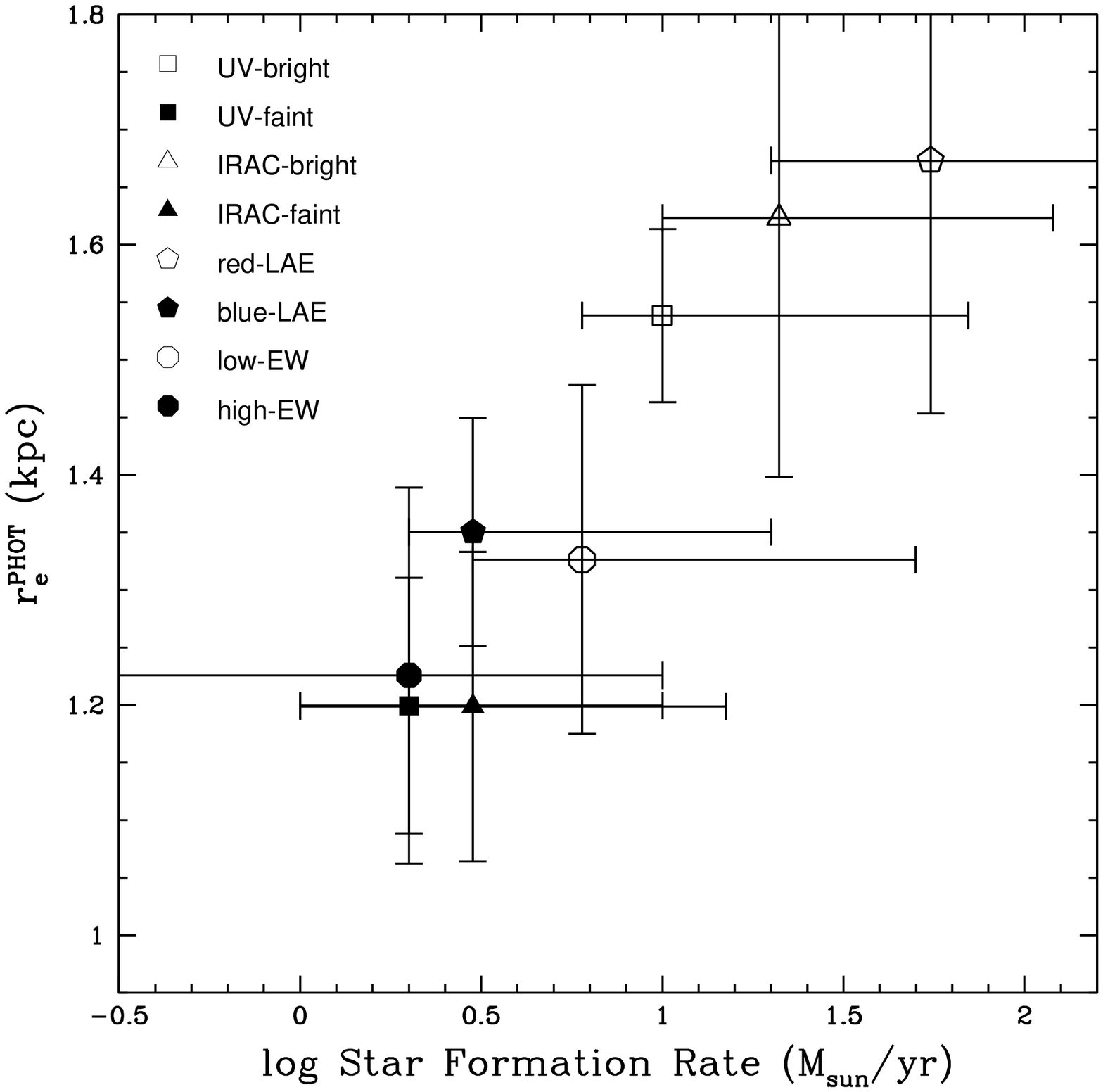}
\caption{Same as Figure~\ref{fig:RvMass}, but plotting fixed-aperture half-light radius as a function of star formation rate.
\label{fig:RvSFR}}
\end{figure}

There is also evidence that the LAE population is becoming more heterogeneous with time.  \citet{Nilsson11} found increased variety in the SEDs of LAEs at $z=2.25$ as compared to higher-redshift samples, with only $15$\% of the galaxies being consistent with a single, young population of stars.  Furthermore, Gu10 find a bimodality in the B$-$R colors of the UV-bright ($R<25$) objects in their sample at $z=2.1$, possibly indicating the presence of a sub-population of LAEs for which a significant quantity of dust is present.  The evidence for heterogeneity in the z2Guaita sample was further demonstrated in \citet{z2SED}, where they looked at the SED properties of a series of subsamples selected using various photometric cuts (see also, Section~\ref{subsec:medstat}).  In Figures~\ref{fig:RvMass}, \ref{fig:RvDust}, and \ref{fig:RvSFR}, we show the relationship between the derived SED properties and the median half-light radii of the z2Guaita subsamples\footnote[3]{Note that the plotted points are not all independent, as there is overlap in the subsamples}.  In all cases, we see a wide range of SED properties, as well as a clear correlation between the LAE size and the properties derived from its SED.  LAEs are found to be larger for galaxies with higher stellar mass, higher dust obscuration, and higher star formation rate \citep[averaged over $100$~Myr, see][]{z2SED}.  These results are broadly consistent with the numerical simulations of \citet{SU10}, who predict that size, stellar mass, and star formation rate will correlate positively with the mass-weighted age of LAEs.  

The apparent heterogeneity of LAEs at z$\sim2$ might indicate the presence of a sub-population of LAEs that are more massive and more evolved.  Such objects would normally have their Ly$\alpha$ emission extinguished by dust, but if the galaxies are accreting a relatively pristine subhalo, the Ly$\alpha$ emission may be originating from a dust-free star-forming region sufficiently separated from the parent halo that the line emission could escape \citep{SU10}.  Alternatively, the dust in these objects may be concentrated in dense clumps, with the Ly$\alpha$ photons resonantly scattering off of the clump surface and escaping the galaxy \citep[e.g.,][]{Neufeld91,HO06,Finkelstein09b}.  In the simulations of \citet{SU10}, only $\sim 30$\% of LAEs at $z=3.1$ are evolved galaxies experiencing delayed accretion of a subhalo onto a parent halo rather than galaxies undergoing their first major burst of star formation.  By $z=2.1$, however, their models show the majority of LAEs ($\sim 70$\%) are in the former category.  This is qualitatively consistent with the increase in size and the broader range of morphological properties that we see at $z=2.1$, as well as the increase in dust reddening seen by \citet{z2SED}.

\begin{deluxetable*}{lcccc}
\tablecaption{Sizes of z=3.1 LAE Subsamples\label{tab:Subsamplesz3}}
\tablewidth{0pt}
\tablehead{
\colhead{Subsample}
&\colhead{Selection}
&\colhead{Number of LAEs\tablenotemark{a}}
&\colhead{Median Size\tablenotemark{b}}
&\colhead{Size Spread\tablenotemark{c}}
\\
&
&
&\colhead{(kpc)}
&\colhead{(kpc)}
}
\startdata
UV-faint &$R>26.3$ &$80$ &$0.97 \pm 0.06$ &$0.55 \pm 0.07$  \\
UV-bright &$R<26.3$ &$56$ &$1.09 \pm 0.09$ &$0.75 \pm 0.12$  \\
IRAC-faint &$f_{3.6}<0.3$~$\mu$~Jy$$ &$31$ &$1.24 \pm 0.14$ &$0.65 \pm 0.08$  \\
IRAC-bright &$f_{3.6}>0.3$~$\mu$~Jy$$ &$11$ &$1.24 \pm 0.24$ &$0.73 \pm 0.23$  \\
low-EW &$EW<66$~\AA$$ &$85$ &$1.06 \pm 0.08$ &$0.67 \pm 0.10$  \\
high-EW &$EW>66$~\AA$$ &$51$ &$0.98 \pm 0.06$ &$0.59 \pm 0.11$  \\
red-LAE &$V-R>0.3$ &$13$ &$1.67 \pm 0.36$ &$1.40 \pm 0.30$  \\
blue-LAE &$V-R<0.3$ &$123$ &$1.00 \pm 0.05$ &$0.58 \pm 0.07$  \\
\enddata

\tablenotetext{a}{Includes only LAEs with {\it HST}/ACS coverage}
\tablenotetext{b}{Median fixed-aperture half-light radius within $0\farcs6$}
\tablenotetext{c}{Interquartile range of fixed-aperture half-light radius}

\end{deluxetable*}

More work needs to be done before we can properly quantify the size evolution of LAEs through cosmic time.  Optimally, we would like to measure size evolution at $z\lesssim1.5$ and $z\gtrsim4$ in order to determine whether the evolution seen here constitutes a sudden change in the LAE population or a more gradual evolution with redshift.  \citet{Taniguchi09} studied the sizes of a sample of $z=5.7$ LAEs, fitting a PSF-convolved model to a stack of 43 LAEs and found a best-fit half-light radius of $0.76$~kpc.  This can be roughly compared to the median GALFIT-derived size of $z=3.1$ LAEs in z3Gronwall, $r_E^{\mathrm{GF}}=0.70$~kpc \citep{GronwallMorph}, but with the caveat that \citet{Taniguchi09} were working with a sample that was considerably brighter in Ly$\alpha$ \citep[$L_{\mathrm{Ly}\alpha}\gtrsim 7 \times 10^{42}$~erg~s$^{-1}$,][]{Murayama07} and \citet{GronwallMorph} were restricting themselves to individual LAEs with S/N$\gtrsim30$ in the {\it HST}/ACS V$_{606}$ images.

It is possible that LAEs will eventually be found to reflect the approximately $H^{-1}(z)$ size evolution of the overall galaxy population \citep{Ferguson04}, but there are good reasons to think that this will not be the case.  LAEs can appear visible to the observer only so long as the Ly$\alpha$ radiation is able to escape the galaxy without being absorbed by dust.  As such, they likely exist for only a short time after galaxy-scale star formation has turned on and before enough dust is formed to extinguish the Ly$\alpha$.  Simulations suggest that this time period could be shorter than $3 \times 10^8$~years \citep[e.g.,][]{MU06}.  If so, then LAEs are only a single snapshot in the history of a galaxy's formation and their size evolution should be much less steep than that of the overall galaxy population.  If, on the other hand, Ly$\alpha$ emission is able to escape from a large fraction of galaxies with previous generations of stars already in place, they should more closely trace the \citet{Ferguson04} law.  The heterogeneity seen in the present samples suggests that the LAE population contains both types of objects, but more work is needed to elucidate whether this division is an actual bimodality or simply a continuous range of properties.  The $z=2.1$ LAE sample, in particular, may contain up to $\sim 15$\% contamination from low-redshift galaxies (see Section~\ref{subsec:classify}) - spectroscopic follow-up is needed to accurately estimate the contamination fraction and isolate the types of objects that contribute to the contamination.  Furthermore, analysis of the deep rest-frame optical (observed-NIR) imaging obtained as part of the Wide Field Camera 3 Early Release Science \citep{Windhorst10} and the Cosmic Assembly Near-infrared Deep Extragalactic Legacy Survey would allow us to determine where most of the stellar mass in these objects lies and whether it coincides spatially with the rest-UV emission from the young stars.


\acknowledgments

This material is based on work supported by NASA through grant number HST-AR-11253.01-A from the Space Telescope Science Institute, which is operated by AURA, Inc., under NASA contract NAS 5-26555, an award issued by JPL/Caltech, by the National Science Foundation under grants AST-0807570 and AST-0807885, and by the Department of Energy under grants DE-FG02-08ER41560 and DE-FG02-08ER41561.  We thank Martin Altmann for the use of his sample of stars in the MUSYC/ECDF-S field.

\bibliographystyle{apj}                       

\bibliography{apj-jour,Bond0414}

\renewcommand{\thefootnote}{\alph{footnote}}

\clearpage

\begin{figure}[t]
\figurenum{1}
\plotone{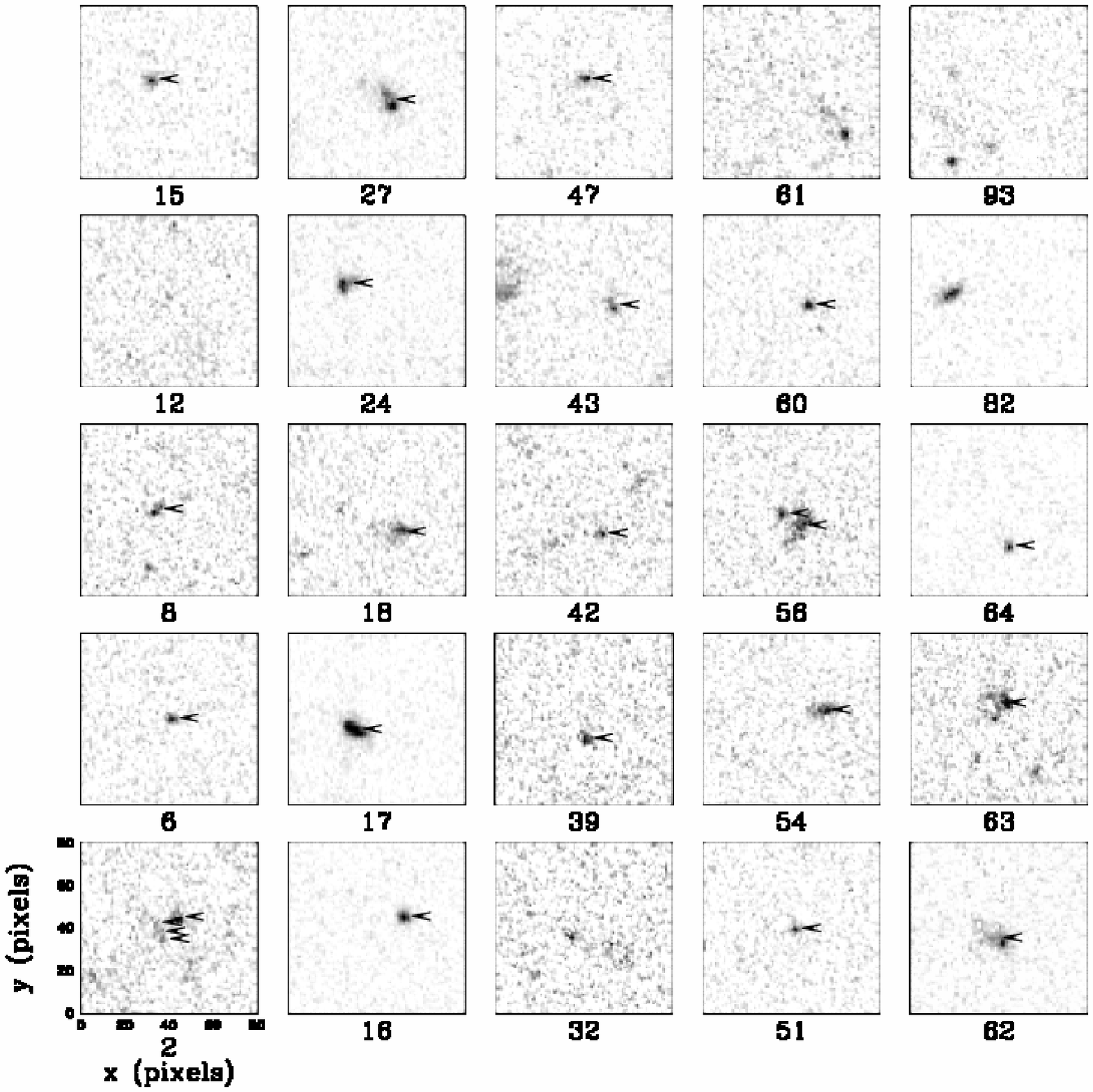}
\caption{Cutouts of $z=2.1$ LAEs extracted from
the GEMS survey images.  We mark components (SExtractor detections within $0\farcs65$ of the cutout center) with arrows.  Numbers underneath the panels are
the corresponding LAE indices from Gu10.
\label{fig:GEMSPanels}}
\end{figure}

\begin{figure}[t]
\figurenum{1}
\plotone{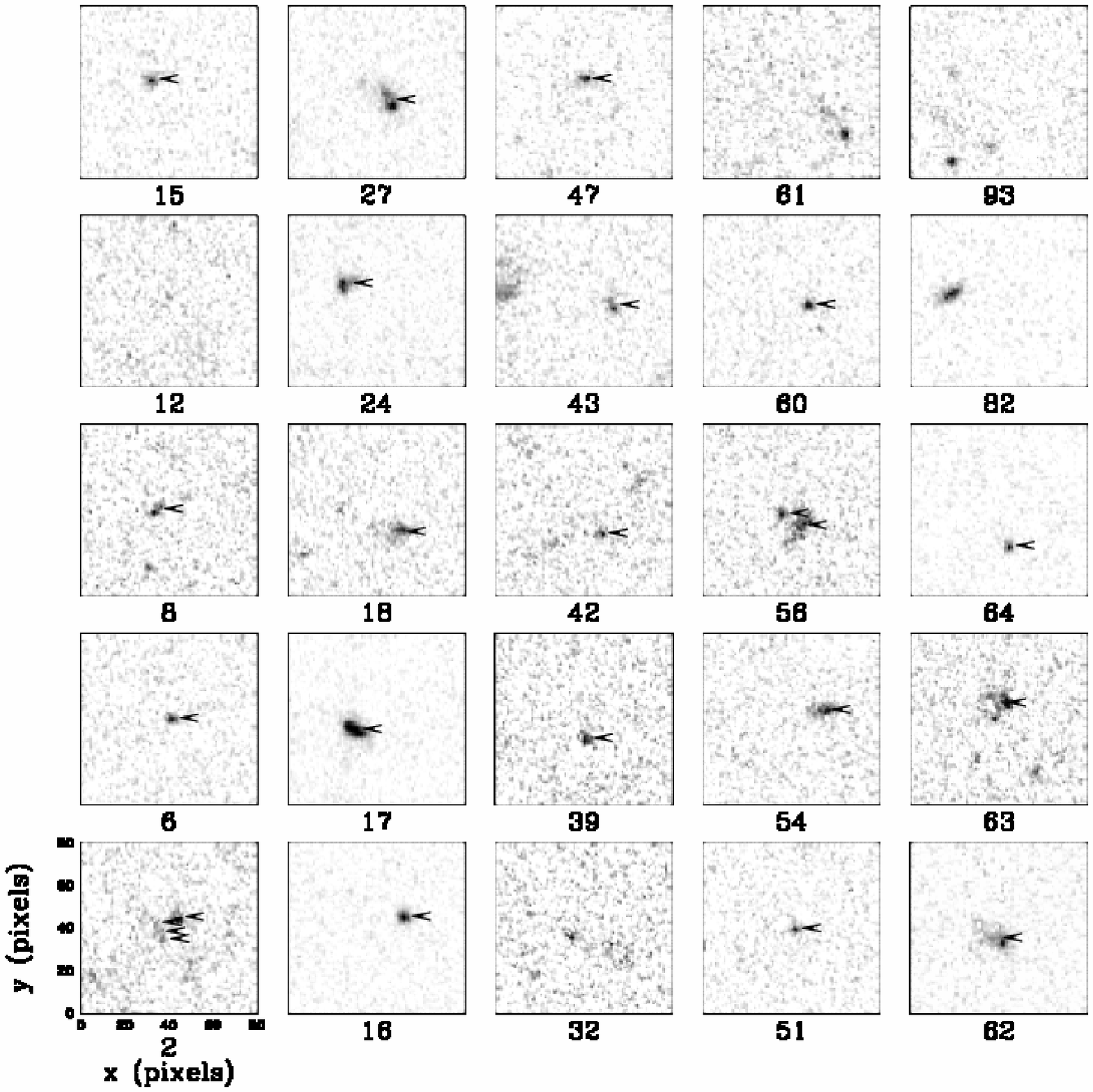}
\caption{(cont.)}
\end{figure}

\begin{figure}[t]
\figurenum{1}
\plotone{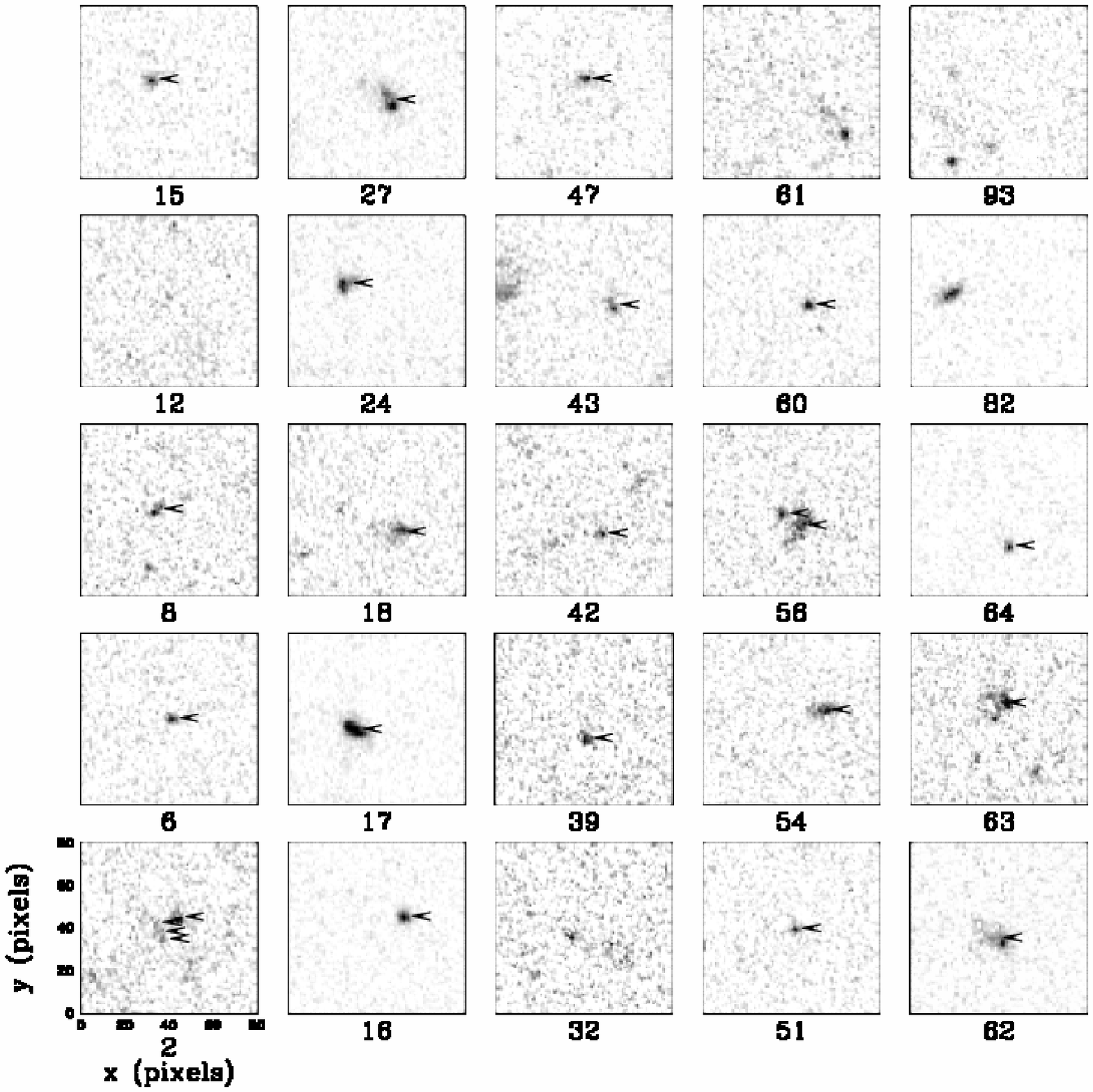}
\caption{(cont.)}
\end{figure}

\begin{figure}[t]
\figurenum{1}
\plotone{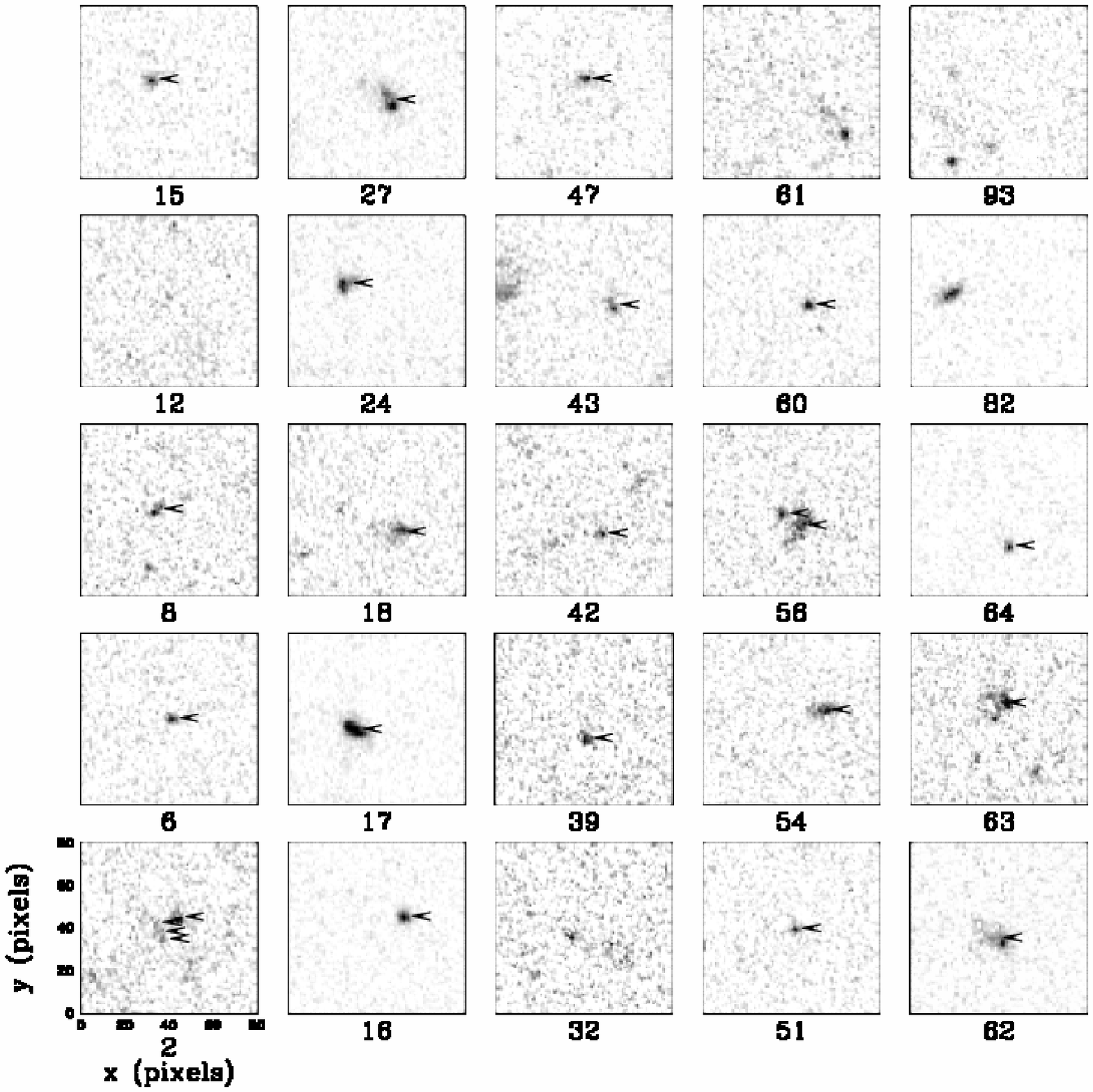}
\caption{(cont.)}
\end{figure}

\begin{figure}[t]
\figurenum{2}
\plotone{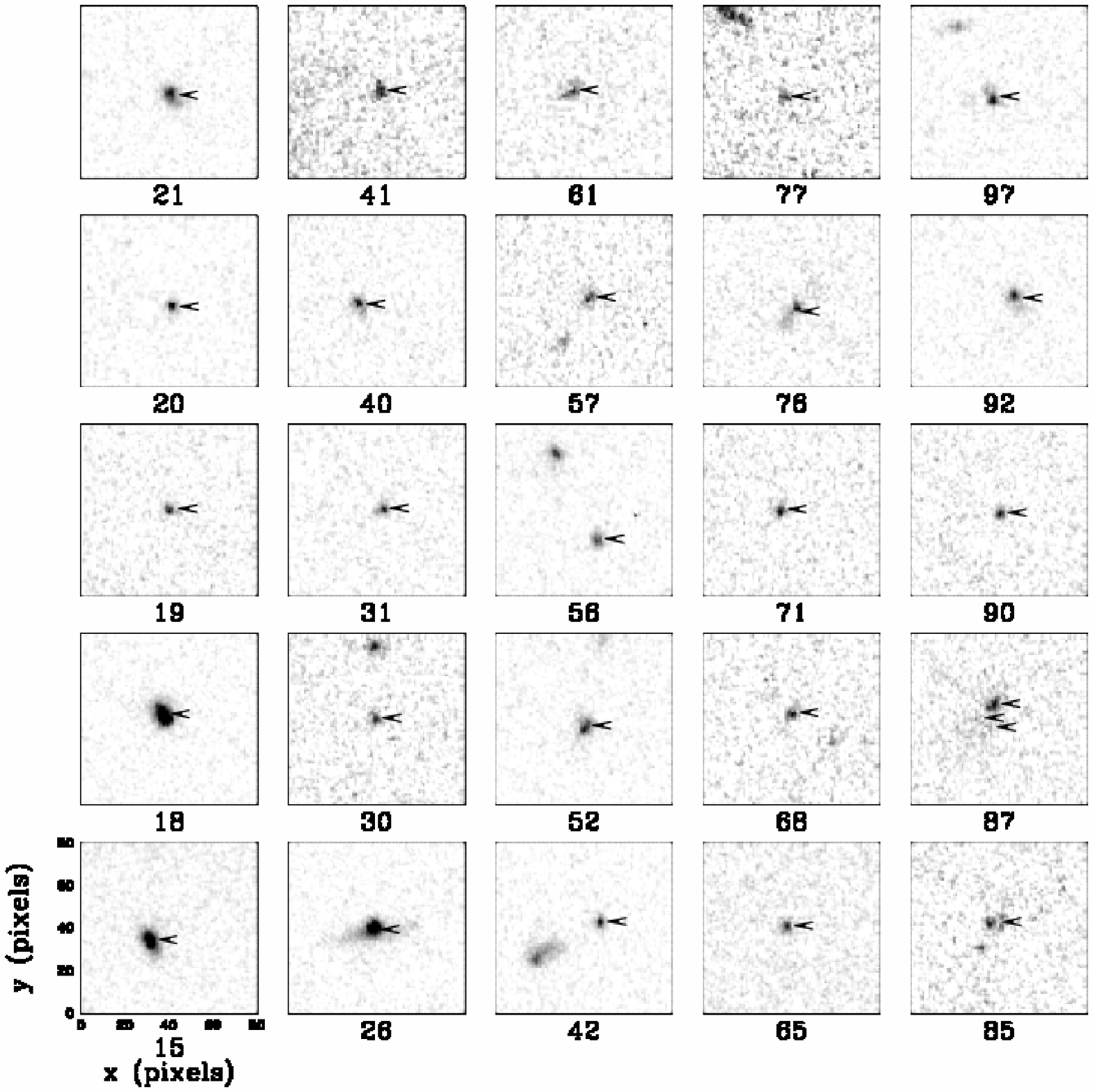}
\caption{Cutouts of $z=3.1$ LAEs extracted from
the GEMS survey images.  The format is the same as in Figure~\ref{fig:GEMSPanels}, except that numbers underneath
the panels correspond to LAE indices from \citet{z2LF}. This sample supplements the $z=3.1$ sample presented in \citet{GronwallLAE} and B09.
\label{fig:GEMSPanelsKO3}}
\end{figure}

\begin{figure}[t]
\figurenum{2}
\plotone{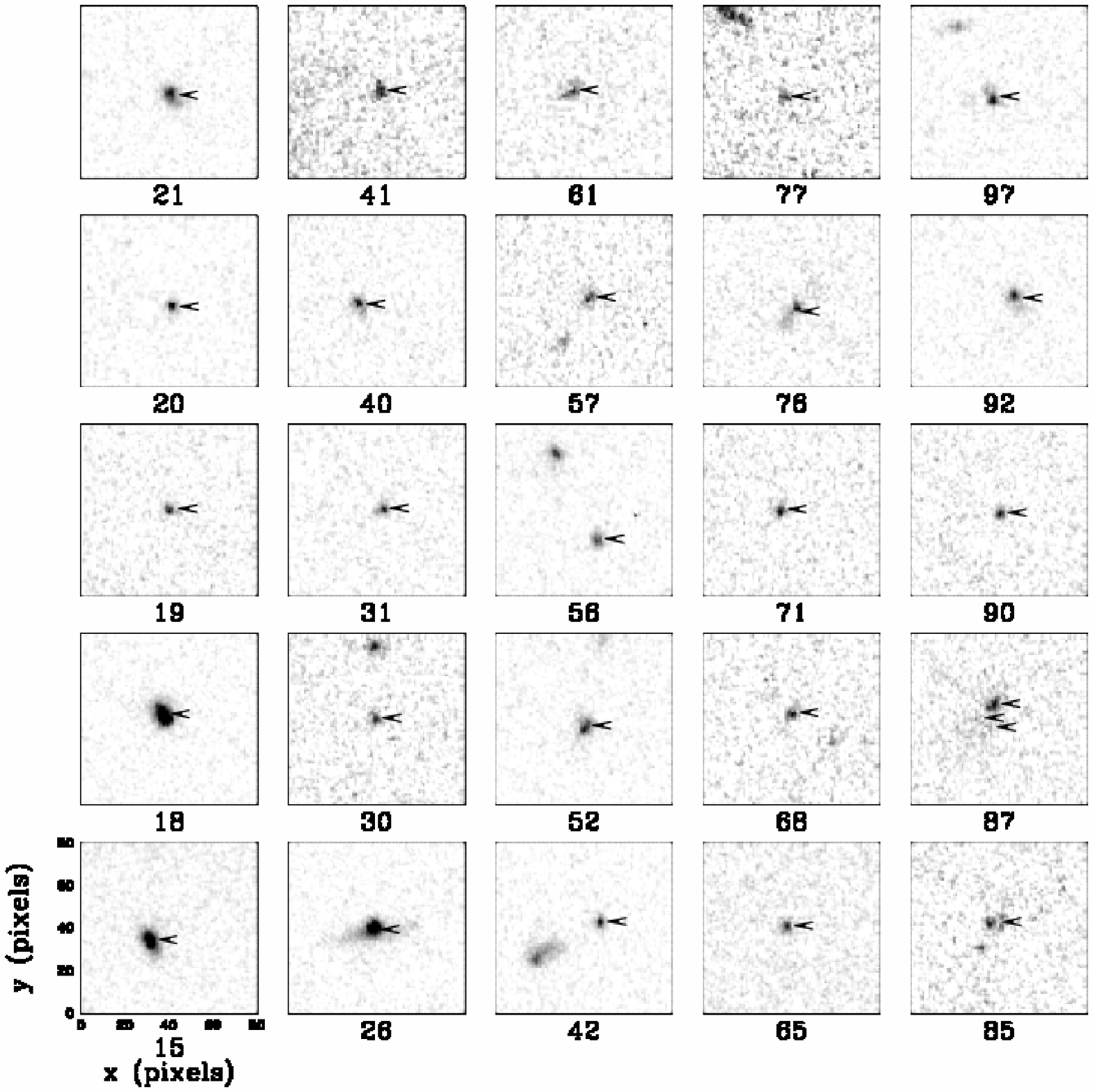}
\caption{(cont.)}
\end{figure}

\begin{figure}[t]
\figurenum{3}
\plotone{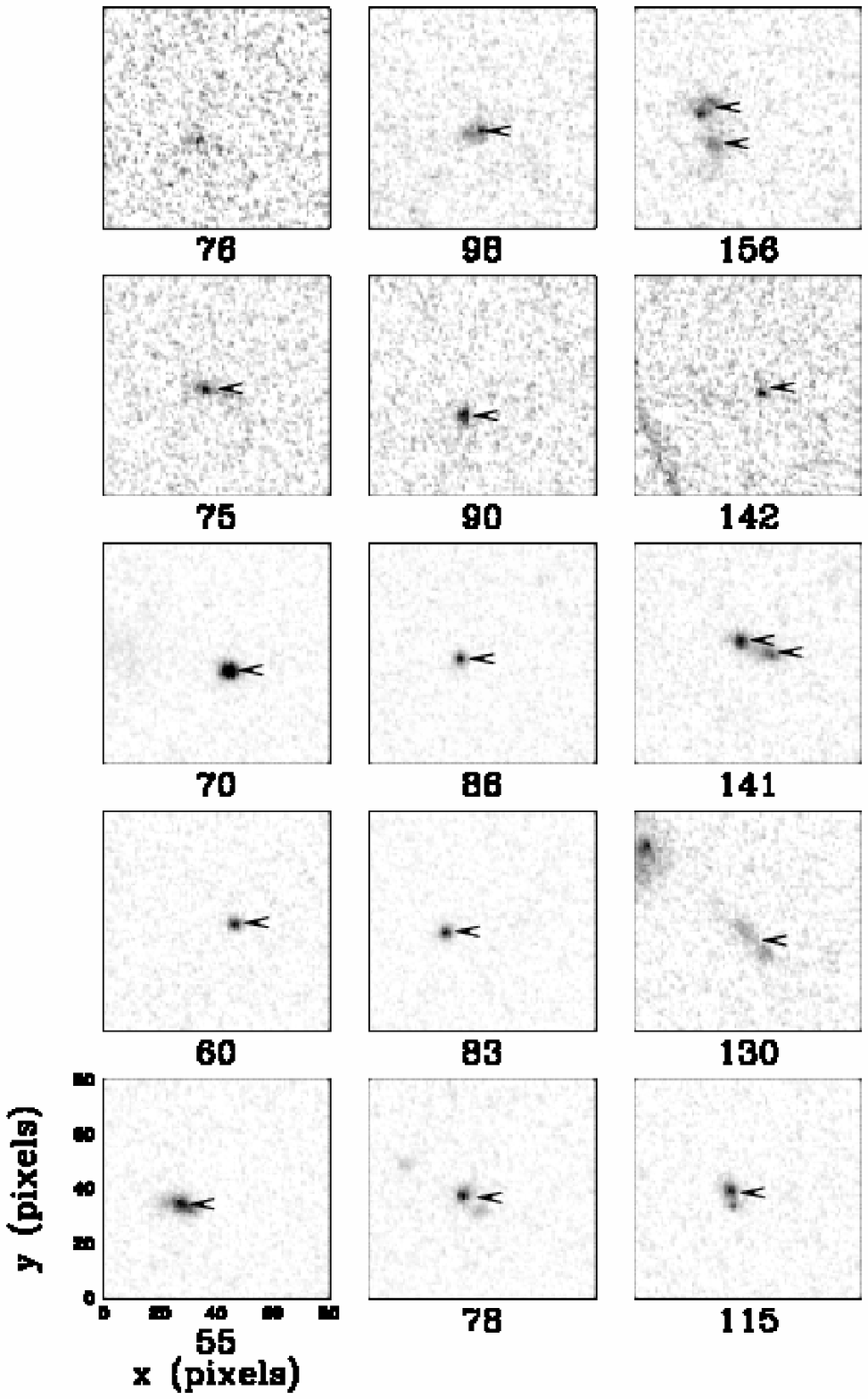}
\caption{Same as Figure~\ref{fig:GEMSPanels}, but for LAEs covered by the GOODS V$_{606}$-band images.
\label{fig:GOODSPanels}}
\end{figure}

\begin{figure}[t]
\figurenum{4}
\plotone{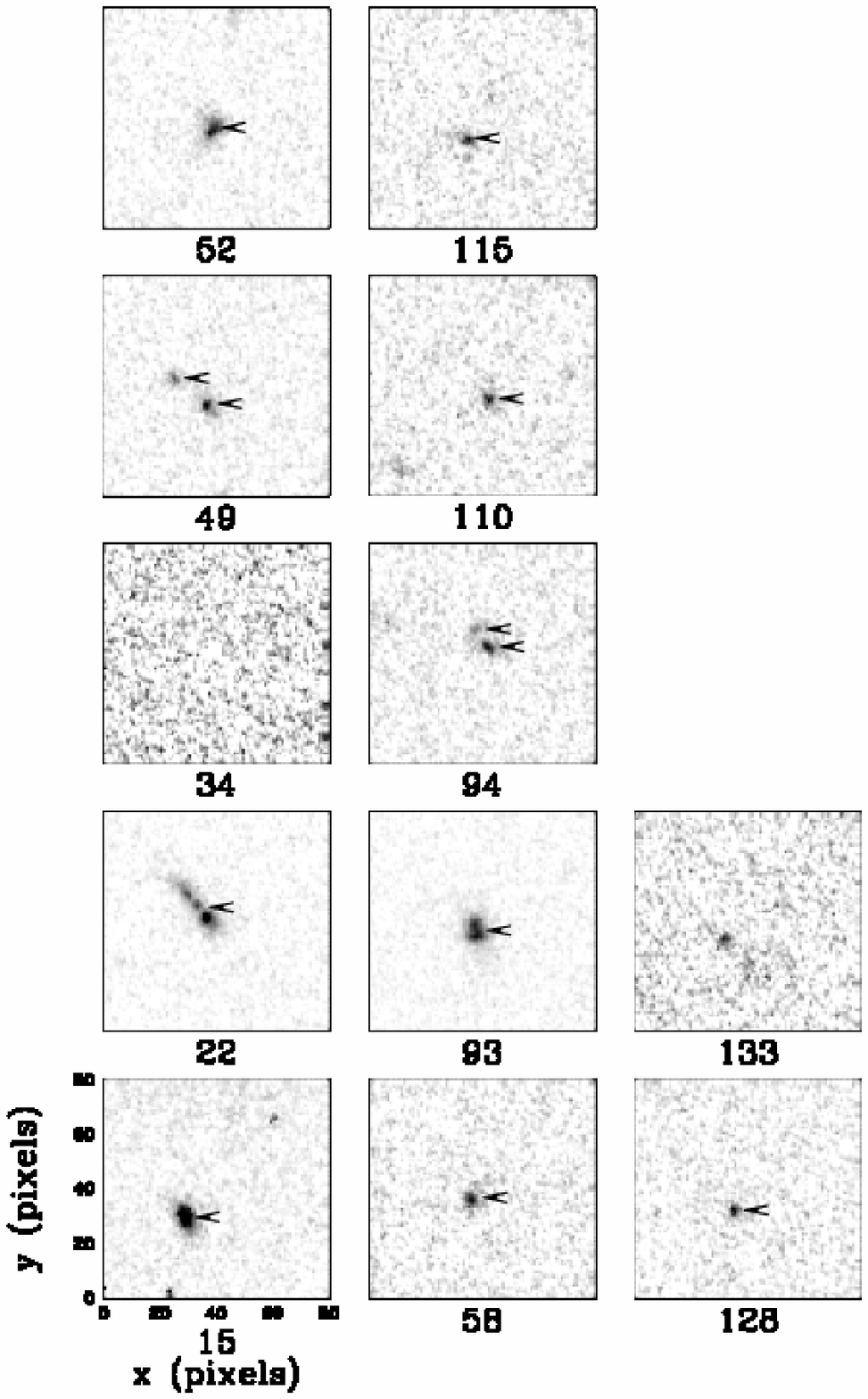}
\caption{Same as Figure~\ref{fig:GEMSPanelsKO3}, but for LAEs covered by the GOODS V$_{606}$-band images.
\label{fig:GOODSPanelsKO3}}
\end{figure}

\begin{figure}[t]
\figurenum{5}
\plotone{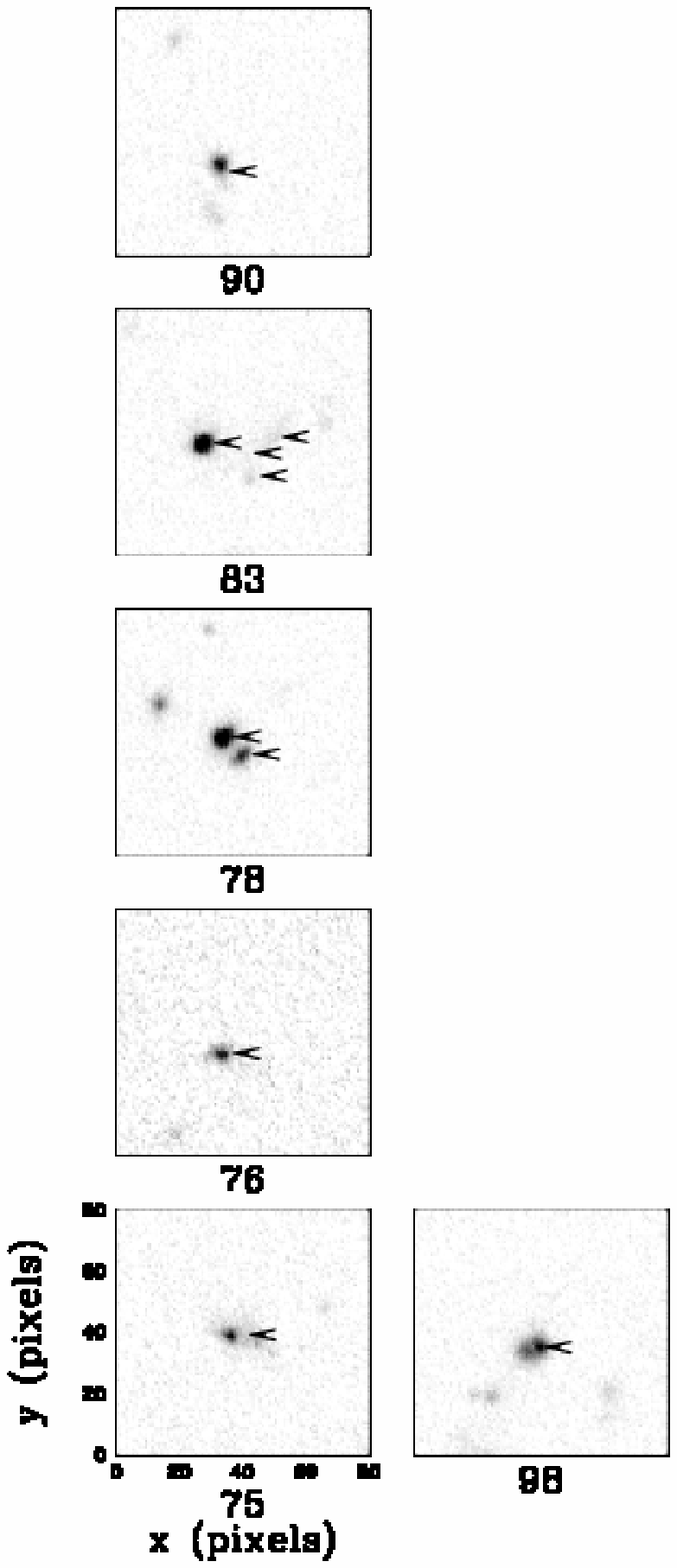}
\caption{Same as Figure~\ref{fig:GEMSPanels}, but for LAEs covered by the HUDF V$_{606}$-band images.
\label{fig:HUDFPanels}}
\end{figure}

\end{document}